\documentclass[12pt,a4paper]{article}

\usepackage{ifthen} 
\newboolean{pdflatex}
\setboolean{pdflatex}{true} 

\newboolean{articletitles}
\setboolean{articletitles}{true} 

\newboolean{uprightparticles}
\setboolean{uprightparticles}{false} 

\def\paperauthors{LHCb collaboration} 
\def\paperasciititle{A method for luminosity determination based on real-time hit reconstruction with the LHCb silicon pixel detector} 
\def\papertitle{A method for luminosity determination based on real-time hit reconstruction with the LHCb silicon pixel detector} 
\def\paperkeywords{{High Energy Physics}, {LHCb}} 
\def\papercopyright{\the\year\ CERN for the benefit of the LHCb collaboration} 
\def\paperlicence{CC BY 4.0 licence}
\def\paperlicenceurl{https://creativecommons.org/licenses/by/4.0/}

\newif\ifEnableSectionTOCLinks
\EnableSectionTOCLinksfalse 

\usepackage[top=1in, bottom=1.25in, left=1in, right=1in]{geometry}

\usepackage{microtype}
\usepackage{lineno}  
\usepackage{xspace} 
\usepackage{caption} 

\usepackage{graphicx}  
\usepackage{color}
\usepackage{colortbl}
\graphicspath{{./figs/}} 

\usepackage{amsmath} 
\usepackage{amssymb}
\usepackage{amsfonts}
\usepackage{upgreek} 

\newcommand*\patchAmsMathEnvironmentForLineno[1]{%
\expandafter\let\csname old#1\expandafter\endcsname\csname #1\endcsname
\expandafter\let\csname oldend#1\expandafter\endcsname\csname
end#1\endcsname
 \renewenvironment{#1}%
   {\linenomath\csname old#1\endcsname}%
   {\csname oldend#1\endcsname\endlinenomath}%
}
\newcommand*\patchBothAmsMathEnvironmentsForLineno[1]{%
  \patchAmsMathEnvironmentForLineno{#1}%
  \patchAmsMathEnvironmentForLineno{#1*}%
}
\AtBeginDocument{%
\patchBothAmsMathEnvironmentsForLineno{equation}%
\patchBothAmsMathEnvironmentsForLineno{align}%
\patchBothAmsMathEnvironmentsForLineno{flalign}%
\patchBothAmsMathEnvironmentsForLineno{alignat}%
\patchBothAmsMathEnvironmentsForLineno{gather}%
\patchBothAmsMathEnvironmentsForLineno{multline}%
\patchBothAmsMathEnvironmentsForLineno{eqnarray}%
}

\usepackage[pdftex,
            pdfauthor={\paperauthors},
            pdftitle={\paperasciititle},
            pdfkeywords={\paperkeywords}]{hyperref}
\usepackage{hyperxmp}
\hypersetup{
    pdfcopyright={Copyright (C) \papercopyright},
    pdflicenseurl={\paperlicenceurl}
}

\usepackage[colorinlistoftodos,textsize=scriptsize]{todonotes}

\usepackage[bottom,flushmargin,hang,multiple]{footmisc}

\usepackage[all]{hypcap} 

\usepackage{xspace}
\usepackage{upgreek}

\def\MagUp {\mbox{\em Mag\kern -0.05em Up}\xspace}

\ifthenelse{\boolean{uprightparticles}}%
{

 \def\PDelta      {\ensuremath{\Delta}\xspace}
 \def\PXi         {\ensuremath{\Xi}\xspace}
 \def\PLambda     {\ensuremath{\Lambda}\xspace}
 \def\PSigma      {\ensuremath{\Sigma}\xspace}
 \def\POmega      {\ensuremath{\Omega}\xspace}
 \def\PUpsilon    {\ensuremath{\Upsilon}\xspace}
 \let\oldPi\Pi
 \def\PPi         {\ensuremath{\oldPi}\xspace}

 \def\PB      {\ensuremath{\mathrm{B}}\xspace}
 \def\PD      {\ensuremath{\mathrm{D}}\xspace}

 \def\PK      {\ensuremath{\mathrm{K}}\xspace}
 \def\Pp      {\ensuremath{\mathrm{p}}\xspace}

 \def\Ps      {\ensuremath{\mathrm{s}}\xspace}

 \def\thebaroffset{0.0em}
}
{

 \mathchardef\PDelta="7101
 \mathchardef\PXi="7104
 \mathchardef\PLambda="7103
 \mathchardef\PSigma="7106
 \mathchardef\POmega="710A
 \mathchardef\PUpsilon="7107
 \mathchardef\PPi="7105
 \def\PB      {\ensuremath{B}\xspace}
 \def\PD      {\ensuremath{D}\xspace}

 \def\PK      {\ensuremath{K}\xspace}
 \def\Pp      {\ensuremath{p}\xspace}

 \def\Ps      {\ensuremath{s}\xspace}

 \def\thebaroffset{0.18em}
}
\newcommand{\offsetoverline}[2][\thebaroffset]{\kern #1\overline{\kern -#1 #2}}%

\makeatletter
\ifcase \@ptsize \relax
  \newcommand{\miniscule}{\@setfontsize\miniscule{4}{5}}
\or
  \newcommand{\miniscule}{\@setfontsize\miniscule{5}{6}}
\or
  \newcommand{\miniscule}{\@setfontsize\miniscule{5}{6}}
\fi
\makeatother

\DeclareRobustCommand{\optbar}[1]{\shortstack{{\miniscule (\rule[.5ex]{1.25em}{.18mm})}
  \\ [-.7ex] $#1$}}

\def\squark    {{\ensuremath{\Ps}}\xspace}

\def\KorKbar {\kern \thebaroffset\optbar{\kern -\thebaroffset \PK}{}\xspace}

\def\D       {{\ensuremath{\PD}}\xspace}

\def\DorDbar {\kern \thebaroffset\optbar{\kern -\thebaroffset \PD}\xspace}

\def\Dp      {{\ensuremath{\D^+}}\xspace}
\def\Dm      {{\ensuremath{\D^-}}\xspace}

\def\DpDm    {\ensuremath{\Dp {\kern -0.16em \Dm}}\xspace}

\def\B       {{\ensuremath{\PB}}\xspace}

\def\BorBbar {\kern \thebaroffset\optbar{\kern -\thebaroffset \PB}\xspace}

\def\Bd      {{\ensuremath{\B^0}}\xspace}

\def\BdorBdbar {\kern \thebaroffset\optbar{\kern -\thebaroffset \Bd}\xspace}

\def\Bs      {{\ensuremath{\B^0_\squark}}\xspace}

\def\BsorBsbar {\kern \thebaroffset\optbar{\kern -\thebaroffset \Bs}\xspace}

\def\Y#1S{\ensuremath{\PUpsilon{(#1S)}}\xspace}

\def\proton      {{\ensuremath{\Pp}}\xspace}

\def\LorLbar     {\kern \thebaroffset\optbar{\kern -\thebaroffset \PLambda}\xspace}

\def\CP                {{\ensuremath{C\!P}}\xspace}

\def\AT#1     {\ensuremath{A_{\mathrm{T}}^{#1}}\xspace}           

\def\C#1      {\ensuremath{\mathcal{C}_{#1}}\xspace}                       
\def\Cp#1     {\ensuremath{\mathcal{C}_{#1}^{'}}\xspace}                    
\def\Ceff#1   {\ensuremath{\mathcal{C}_{#1}^{\mathrm{(eff)}}}\xspace}        
\def\Cpeff#1  {\ensuremath{\mathcal{C}_{#1}^{'\mathrm{(eff)}}}\xspace}       
\def\Ope#1    {\ensuremath{\mathcal{O}_{#1}}\xspace}                       
\def\Opep#1   {\ensuremath{\mathcal{O}_{#1}^{'}}\xspace}                    

\newcommand{\nospaceunit}[1]{\ensuremath{\text{#1}}}
\newcommand{\aunit}[1]{\ensuremath{\text{\,#1}}}

\newcommand{\tev}{\aunit{Te\kern -0.1em V}\xspace}
\newcommand{\gev}{\aunit{Ge\kern -0.1em V}\xspace}
\newcommand{\mev}{\aunit{Me\kern -0.1em V}\xspace}
\newcommand{\kev}{\aunit{ke\kern -0.1em V}\xspace}
\newcommand{\ev}{\aunit{e\kern -0.1em V}\xspace}

\newcommand{\mevc}{\ensuremath{\aunit{Me\kern -0.1em V\!/}c}\xspace}
\newcommand{\gevc}{\ensuremath{\aunit{Ge\kern -0.1em V\!/}c}\xspace}
\newcommand{\mevcc}{\ensuremath{\aunit{Me\kern -0.1em V\!/}c^2}\xspace}
\newcommand{\gevcc}{\ensuremath{\aunit{Ge\kern -0.1em V\!/}c^2}\xspace}

\def\cm   {\aunit{cm}\xspace}

\def\mm   {\aunit{mm}\xspace}

\def\mum  {\ensuremath{\,\upmu\nospaceunit{m}}\xspace}

\def\mub{\ensuremath{\,\upmu\nospaceunit{b}}\xspace}
\def\sec  {\ensuremath{\aunit{s}}\xspace}
\def\ms   {\ensuremath{\aunit{ms}}\xspace}

\def\mhz  {\ensuremath{\aunit{MHz}}\xspace}
\def\khz  {\ensuremath{\aunit{kHz}}\xspace}
\def\hz   {\ensuremath{\aunit{Hz}}\xspace}

\def\gsim{{~\raise.15em\hbox{$>$}\kern-.85em
          \lower.35em\hbox{$\sim$}~}\xspace}
\def\lsim{{~\raise.15em\hbox{$<$}\kern-.85em
          \lower.35em\hbox{$\sim$}~}\xspace}

\def\sqs   {\ensuremath{\protect\sqrt{s}}\xspace}
\def\sqsnn {\ensuremath{\protect\sqrt{s_{\scriptscriptstyle\text{NN}}}}\xspace}

\def\tell1  {TELL1\xspace}
\def\ukl1   {UKL1\xspace}

\newcommand{\eg}{\mbox{\itshape e.g.}\xspace}
\newcommand{\ie}{\mbox{\itshape i.e.}\xspace}

\newcommand{\vs}{\mbox{\itshape vs.}\xspace}

\newcommand{\lhcborcid}[1]{\href{https://orcid.org/#1}{\hspace*{0.1em}\raisebox{-0.45ex}{\includegraphics[width=1em]{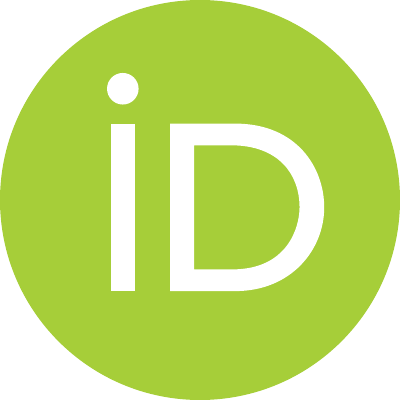}}}}

\hypersetup{
  colorlinks   = true, 
  urlcolor     = blue, 
  linkcolor    = blue, 
  citecolor    = red   
}

\ifEnableSectionTOCLinks
    \usepackage[explicit]{titlesec} 
    
    \let\oldcontentsline\contentsline
    \renewcommand

    \titleformat{\section}{\normalfont\Large\bf}{\hyperlink{tocsection.\thesection}{{\thesection} \parbox[t]{\dimexpr\textwidth-1pc}{#1}}}{1pc}{}

    \titleformat{\subsection}{\normalfont\bf}{\hyperlink{tocsubsection.\thesubsection}{{\thesubsection} \parbox[t]{\dimexpr\textwidth-1pc}{#1}}}{1pc}{}

    \titleformat{name=\section,numberless}[display]{}{}{0pt}{\normalfont\Huge\bfseries #1}
\fi

\usepackage{cite} 
\usepackage{mciteplus}

\usepackage{longtable} 

\usepackage{xcolor}
\usepackage{lineno}
\usepackage{siunitx}
\usepackage{comment}
\usepackage{xspace}

\renewcommand{\L}{\ensuremath{\mathcal{L}}\xspace}
\newcommand{\muvis}{\ensuremath{\mu_{\text{vis}}}\xspace}
\newcommand{\mupvis}{\ensuremath{{\mu'}_{\text{vis}}}\xspace}
\newcommand{\svis}{\ensuremath{\sigma_{\text{vis}}}\xspace}

\usepackage{./mciteplus}
\mciteErrorOnUnknownfalse

\begin{document}

\renewcommand{\thefootnote}{\fnsymbol{footnote}}
\setcounter{footnote}{1}


\begin{titlepage}
\pagenumbering{roman}

\vspace*{-1.5cm}
\centerline{\large EUROPEAN ORGANIZATION FOR NUCLEAR RESEARCH (CERN)}
\vspace*{1.5cm}
\noindent
\begin{tabular*}{\linewidth}{lc@{\extracolsep{\fill}}r@{\extracolsep{0pt}}}
\ifthenelse{\boolean{pdflatex}}
{\vspace*{-1.5cm}\mbox{\!\!\!\includegraphics[width=.14\textwidth]{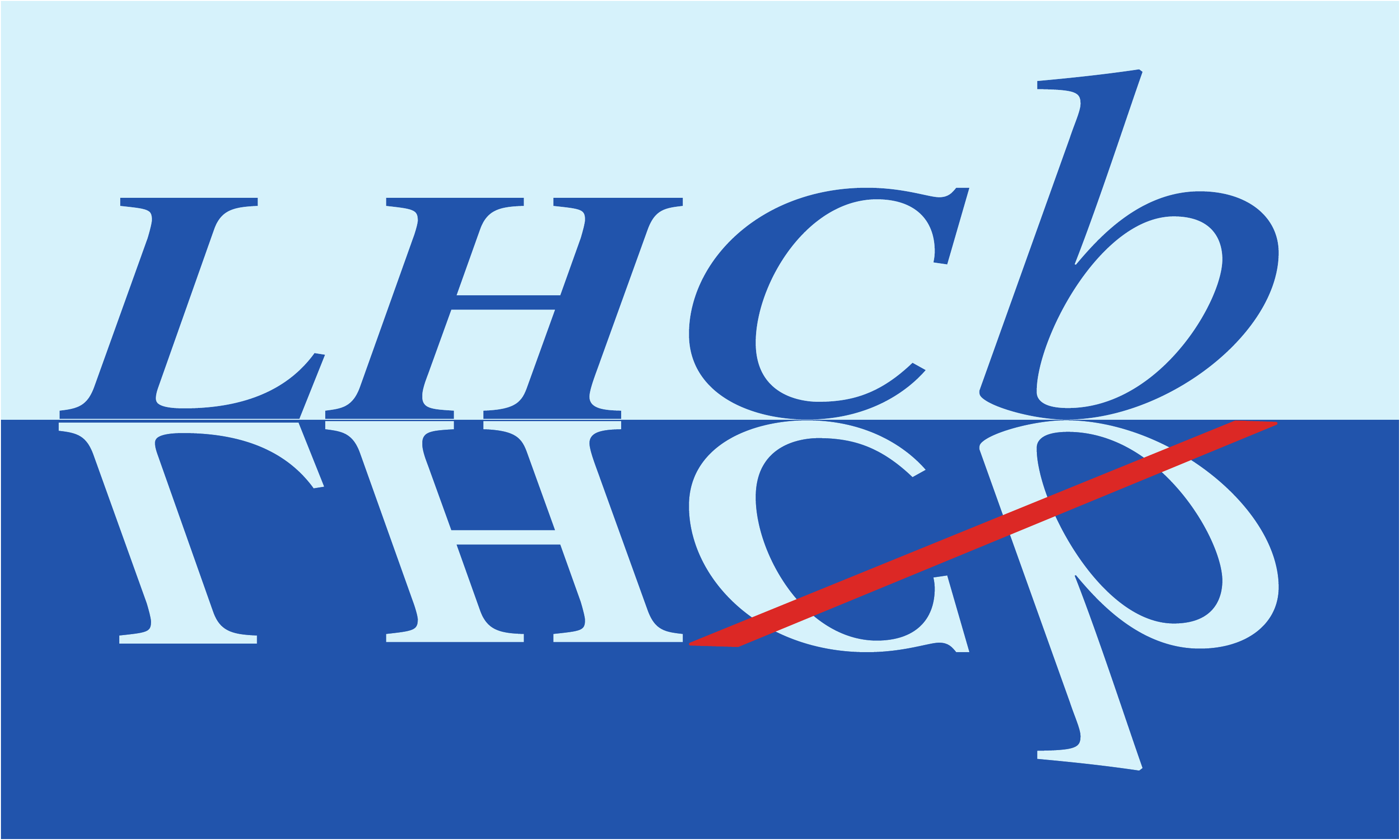}} & &}%
{\vspace*{-1.2cm}\mbox{\!\!\!\includegraphics[width=.12\textwidth]{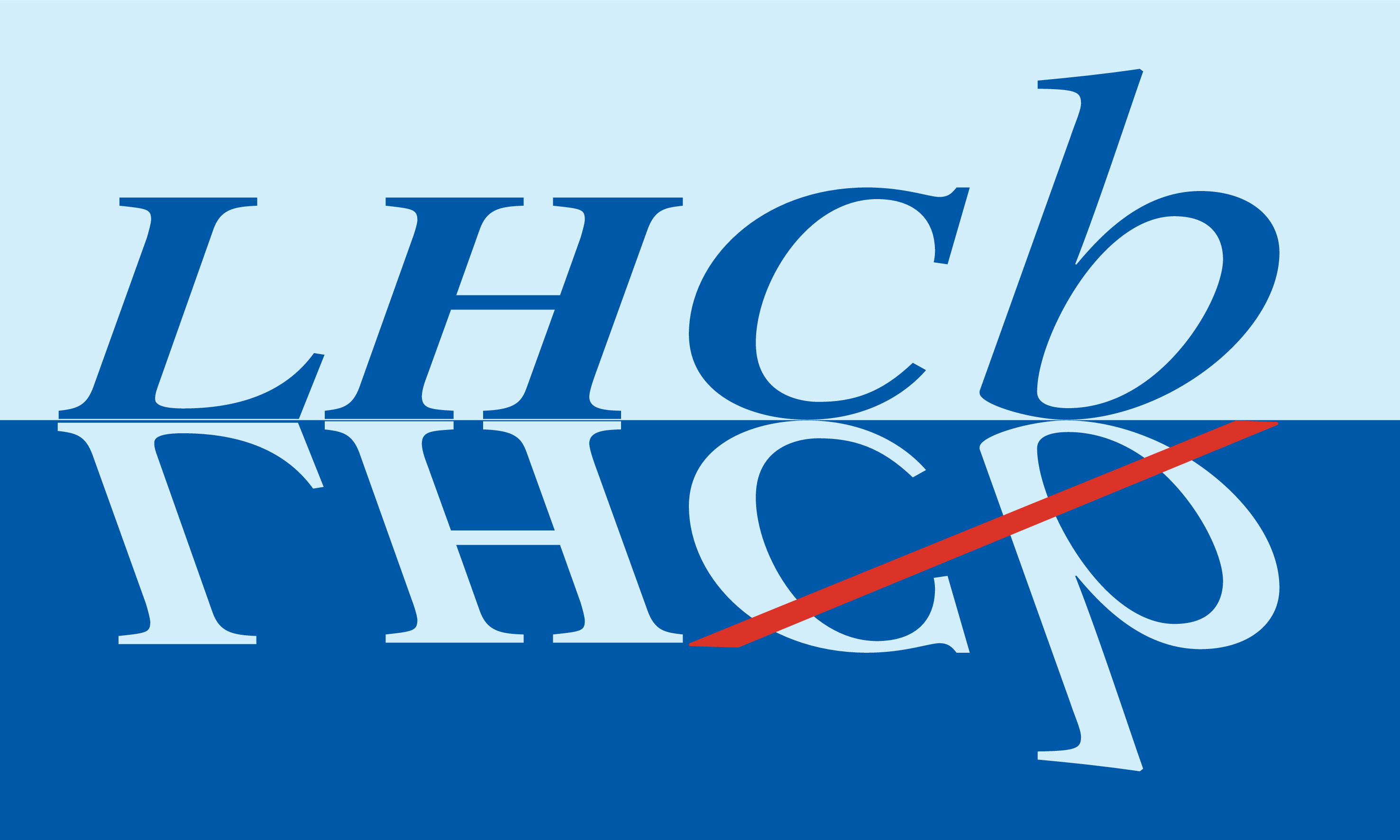}} & &}%
\\
 & & CERN-EP-2026-024 \\  
 & & LHCb-DP-2025-006 \\  
 & & February 15, 2026 \\ 
 & & \\
\end{tabular*}

\vspace*{2.5cm}

{\normalfont\bfseries\boldmath\huge
\begin{center}
  \papertitle 
\end{center}
}

\vspace*{0.5cm}

\begin{center}
\paperauthors\footnote{Authors are listed at the end of this paper.}
\end{center}

\vspace{\fill}

\begin{abstract}
  \noindent
   The data acquisition system of the upgraded LHCb experiment includes the fast reconstruction of all hits in the vertex locator (VELO) pixel detector at the beam-crossing rate of 40\mhz, implemented as on-the-fly clustering embedded in the firmware of the readout board FPGAs. The availability of a high rate of reconstructed clusters in real time enables a new fast approach for measuring luminosity and monitoring the LHCb luminous region, directly at the detector readout level.
   This methodology has been implemented as an array of real-time cluster counters in the VELO readout FPGAs and has been in operation since the start of the 2024 physics run of LHCb. This paper describes the methodology and its features and performance, both on proton-proton and lead-lead collision data.
   The method shows a statistical resolution better than the percent level, and a sensitivity to variable running conditions of the same level. This is achieved with an intrinsic time granularity better than 100\ms, undersampled to 3\sec for analysis purposes. Nonlinear behaviour is compatible with zero in a luminosity range including the LHCb Run~3 operating point.
  
\end{abstract}

\vspace*{2.0cm}

\begin{center}
  Submitted to JINST
\end{center}

\vspace{\fill}

{\footnotesize 
\centerline{\copyright~\papercopyright. \href{\paperlicenceurl}{\paperlicence}.}}
\vspace*{2mm}

\end{titlepage}


\newpage
\setcounter{page}{2}
\mbox{~}
%
%
%
%


\renewcommand{\thefootnote}{\arabic{footnote}}
\setcounter{footnote}{0}


\cleardoublepage


\pagestyle{plain} 
\setcounter{page}{1}
\pagenumbering{arabic}



\section{Introduction}
\label{sec:intro}

The LHCb experiment, installed at the Large Hadron Collider (LHC)~\cite{Bruning:2004ej}, was primarily designed for precision tests of \CP violation in  the decays of heavy-flavour hadrons, and for the study of their rare decays. Over time, its physics programme has successfully expanded to encompass a broad range of topics, including quantum chromodynamics, hadron spectroscopy, heavy-ion collisions, and fixed-target physics~\cite{LHCb-DP-2008-001}. 

With respect to other LHC experiments, LHCb is designed to operate at a lower luminosity, kept constant via a procedure known as luminosity levelling~\cite{Follin:2014nva}. A reliable real-time measurement of instantaneous luminosity, with a precision at least as good as 5\%, is therefore essential to ensure correct LHCb operation~\cite{LHCb-TDR-022}. 

In this paper, a new algorithm for the determination of instantaneous luminosity is proposed and characterised. This algorithm relies solely on detector hits, \ie low-level information made available at the very first stage of data acquisition by a new clustering procedure described in Ref.~\cite{Bassi:2023jpv}. It can run continuously and independently of the software-based event reconstruction.\footnote{Throughout this paper, the term \textit{event} indicates a bunch-crossing event.} 

The paper is organised as follows: Sect.~\ref{sec:intro_detector} describes the LHC collider and the relevant features of the LHCb experiment, Sect.~\ref{sec:lumi_intro} introduces the concept of luminosity and the methods normally used to measure it, Sect.~\ref{sec:counters} describes how hit counters are implemented to measure luminosity, and discusses the design choices. Then, Sect.~\ref{sec:calib} explains how these counters are calibrated using collision data, Sect.~\ref{sec:results} discusses the performance of the method and the possible sources of systematic uncertainties. Finally, Sect.~\ref{sec:concl} summarises the findings.

\section{Experimental context}\label{sec:intro_detector}
The LHC is the highest-energy proton-proton (\proton\proton) collider in the world, also capable of colliding heavier nuclei, \eg lead ions (Pb). 
It is a radio-frequency machine where each beam is divided into 3564 bunches. During the 2024 nominal running conditions for \proton\proton collision physics, 
2133 bunches were colliding at the LHCb interaction point, each one containing up to about $1.6\times 10^{11}$ protons~\cite{Bruning:2004ej}.

The LHCb detector underwent a major upgrade in view of the ongoing LHC Run~3~\cite{LHCb-DP-2022-002}, to support a fivefold increase of instantaneous luminosity compared to the previous running periods, \ie a target luminosity of $\mathcal{L}=\SI{2e33}{cm^{-2}s^{-1}}$, corresponding to 2000~\!\hz/\!\mub.
The new detector is able to reconstruct events at a rate of \SI{40}{\mega\hertz} and its finer granularity allows for smooth operation with increased event multiplicity.
The tracking system has been completely replaced. In particular, a new silicon pixel VErtex LOcator (VELO) acts as the tracking detector closest to the \proton\proton interaction region~\cite{LHCb-TDR-013}. The VELO is placed in a vacuum box that locally replaces the LHC vacuum pipe. It consists of 26 detector layers placed at different $z$ positions surrounding the luminous region, where the $z$ axis lies along the beamline and its origin lies within the VELO. Each layer is composed of two modules, which can move horizontally as far as 25\mm and as close as 5.1\mm from the beam axis. This design ensures detector safety during the beam injection phase as well as optimal performance when the beams are set up for physics data taking. Modules are made of 4 hybrid silicon pixel sensors, each read out by 3 ASICs with a  \mbox{$55\mum\,\times\,55\mum$} pitch~\cite{LHCb-TDR-013}.
At the upstream edge of the VELO detector, a new system for fixed-target physics, SMOG2~\cite{LHCb-DP-2024-002}, was installed in the context of the upgrade. It consists of a 20\cm-long aluminium storage cell in which various gases can be injected
to enable the concurrent acquisition of both \proton\proton (PbPb) and \proton-nuclei (Pb-nuclei) collision events.

The readout electronics of all subdetectors have been overhauled for Run~3. Data collected at every bunch crossing are propagated from front-end (FE) to back-end (BE) electronics  through radiation-tolerant optical links. Custom PCI-express BE boards (PCIe40) are mounted in computer servers~\cite{Cachemiche_2016}. These boards contain an Intel Arria-10 GX 1150 FPGA~\cite{arria10ds} whose firmware can be configured either for data acquisition (TELL40) or control (SOL40 and SODIN).
The experiment control system (ECS) allows the user to communicate with several components of the DAQ, including the SOL40, TELL40 and FE boards, the power supplies and the environmental monitors~\cite{LHCb-DP-2022-002, Barbosa:2019vkc}.
The TELL40 boards decode, preprocess and format the data for transmission to the event building computing farm. 

The upgraded LHCb experiment makes use of a triggerless readout architecture~\cite{LHCB-TDR-018,LHCB-TDR-017}.
The entire readout chain, from FE to the processing farm, is controlled by the Timing and Fast Control (TFC) system~\cite{LHCb-DP-2022-002}. The TFC uses the LHC filling scheme to start the DAQ process at all colliding bunch crossings and on a subset of the noncolliding ones.
For each bunch crossing selected by the TFC, data from each sub-detector are assembled into event fragments, and then gathered together for event processing. For each event, detector information is used to create higher-level objects. The process of reconstruction and data filtering happens in two software-based trigger stages, HLT1 and HLT2, where high-level physical quantities such as particle tracks and collision vertices are reconstructed and made available for both real-time and offline analysis~\cite{LHCb-DP-2022-002}.

By design, LHCb operates at a levelled luminosity~\cite{Follin:2014nva}, kept at a lower value with respect to ATLAS and CMS.
To achieve this, the LHC beams colliding at LHCb are steered based on real-time luminosity feedback from the experiment. In Run~3, the LHCb luminosity levelling strategy requires a precision of at least 5\% on the measurement of instantaneous luminosity~\cite{LHCb-TDR-022}.
In the Run~1 and Run~2 data-taking periods, the activity in the LHCb calorimeter was used to measure the instantaneous luminosity, using hence information available at the first stage of the trigger (L0), implemented in hardware.
In Run~3, following the removal of the L0 trigger stage, new methods have been introduced to monitor luminosity. 
A dedicated detector, Probe for LUminosity MEasurement (PLUME), has been designed and installed for this purpose~\cite{LHCb-TDR-022}, and several other quantities, either available at the detector readout level or computed in HLT1, are used as real-time luminometers. A method using hit counting in the muon system has been studied with Run~2 data~\cite{sym14050860}. A metal-foil based detector, the RMS-R3, is used to monitor beam and background conditions continuously during LHCb operation, and, when calibrated on PLUME, it can be used to monitor luminosity in real time~\cite{Pugatch_2025}. The two ring-imaging Cherenkov (RICH) detectors, used for particle identification, have also proved capable of providing real-time luminosity measurements using both MaPMT anode currents and hit counting~\cite{LHCb-DP-2024-003}. The availability of independent luminosity measurements is important in order to maximise the reliability of the luminosity-levelling feedback, as well as to use them for cross-calibration in measurements of  both instantaneous and integrated luminosity. The possibility of using VELO hits reconstructed in real time to measure luminosity~\cite{Passaro:2024fkw} has the additional potential of producing an accurate measurement with a fast time response. This new method has been implemented, and it is described in this paper.
The algorithm, presented in detail in Sect.~\ref{sec:counters}, is developed around two core components. The first one operates on the TELL40 boards, running in parallel with standard data taking and leveraging the new readout architecture to continuously process all collision data. The second one is integrated into the ECS software, where it performs continuous luminosity monitoring based on the data processed at the readout level.

\section{Luminosity determination}
\label{sec:lumi_intro}

The luminosity measurement can be performed by counting the average number of detected occurrences per bunch crossing \muvis of a given reference process~\cite{Grafstrom:2015foa}. Given the visible cross-section \svis of the considered process, the average luminosity at the collision point is expressed as
\begin{align}
    \L = f\, N_{\rm bb} \, \frac{\muvis}{\svis},\label{eqn:lumi}
\end{align}
where $f=11.245~\khz$ is the LHC revolution frequency and $N_{\rm bb}$ the number of colliding bunches. The per-bunch instantaneous luminosity is obtained by measuring \muvis over only one bunch-crossing pair.

Any physics process producing a signal in a detector can provide a suitable reference for luminosity determination. Measuring the rate of such a process allows to monitor the luminosity.
However, the performance of different proxies for luminosity monitoring depends on factors such as noise, resolution, linearity, dynamic range, long-term stability and response time.
A reliable real-time luminosity estimator can be obtained by taking a quantity that is stable in time and proportional to the luminosity itself, and averaging it across all colliding bunches. Taking such an average will be referred to as \textit{average method} in the remainder of this paper. 

Depending on the characteristics of the counter, saturation effects and nonlinearities may arise, especially in conditions of high luminosity, as well as detector instabilities.
One way to mitigate saturation effects is by assuming Poisson statistics, and only counting the number of empty events. The number $k$ of visible interactions per event follows a Poisson distribution with mean \muvis, and thus 
\begin{align}
P\left(k\vert\muvis\right) = \frac{{\left(\muvis\right)}^k e^{-\muvis}}{k!} \;\Longrightarrow \;P\left(0\vert\muvis\right) = e^{-\muvis} \;\Longrightarrow\; \muvis = -\ln P\left(0\right),\label{eqn:log0}
\end{align}
where $P\left(0\right)$ is the fraction of empty events recorded for the given luminosity counter. This method of measuring \muvis, hereinafter referred to as the \textit{log0 method}, also known as \textit{zero-counting} in the literature \cite{Zaitsev:2000js, LHCb-PAPER-2014-047}, is better suited for small-acceptance luminometers, but can become biased and lose statistical precision if the number of empty events becomes too small.

\section{VELO-hit luminosity counters}
\label{sec:counters}

Quantities measured by silicon vertex detectors have been shown to provide precise, reliable and stable luminosity measurements. In Run~1, the best LHCb luminosity monitor was found to be the number of VELO tracks per event~\cite{LHCb-PAPER-2014-047}. Counting the number of reconstructed collision vertices, the total number of VELO hits, or the total number of active pixels are also viable methods.
Similarly, the CMS collaboration has successfully performed a precise measurement of the integrated luminosity using offline-reconstructed pixel-cluster counters from the CMS inner tracker~\cite{CMS:2021xjt}. 

Counting particle hits, reconstructed as clusters of neighbouring pixels, offers a better proxy for luminosity than counting raw pixels, being more directly related to the physics processes 
and more robust with respect to variations of the detector response  due to radiation damage or changes of operational settings.
However, clustering neighbouring active pixels into particle hits may potentially lead to a saturation effect due to the physical overlap of hits on the detector sensor. The capability of individually resolving hits depends on the pixel pitch. Studies based on pseudoexperiments, under the empirically verified assumption that the hits on the VELO sensors are distributed as $r^{-2}$, where $r$ is the radial distance from the beam axis, show that the rate of merged hits is well described by a power-law $f(r)\propto r^{-3}$. At the design Run~3 luminosity, for $r\gtrsim 11\mm$, merged hits on the VELO are less than $0.5\%$. This effect can therefore be made negligible by measuring the luminosity at an appropriate distance from the beam axis.

In the upgraded LHCb detector, with the introduction of real-time clustering on the VELO readout boards~\cite{Hennessy:2021aec}, it becomes possible to look at particle hits  within the readout boards themselves.
A two-dimensional (2D) clustering algorithm embedded on the TELL40 boards makes use of spare FPGA resources not used by the readout firmware~\cite{Bassi:2023jpv}. It produces a list of reconstructed hits, each characterised by the $x,y$ coordinates of its centroid, and by the shape and size of the cluster, as well as quality flags.
The list of reconstructed hits is assembled into event data and transmitted to the event-building server, where raw information from different subdetectors is gathered together before being processed by the HLT1 trigger stage.
Thanks to the triggerless readout, all data are processed at every bunch crossing, making about $ 10^{11}$~hits/\!\sec available to perform measurements in real time, such as luminosity.
\begin{figure}
\centering
\includegraphics[height=4.05cm]{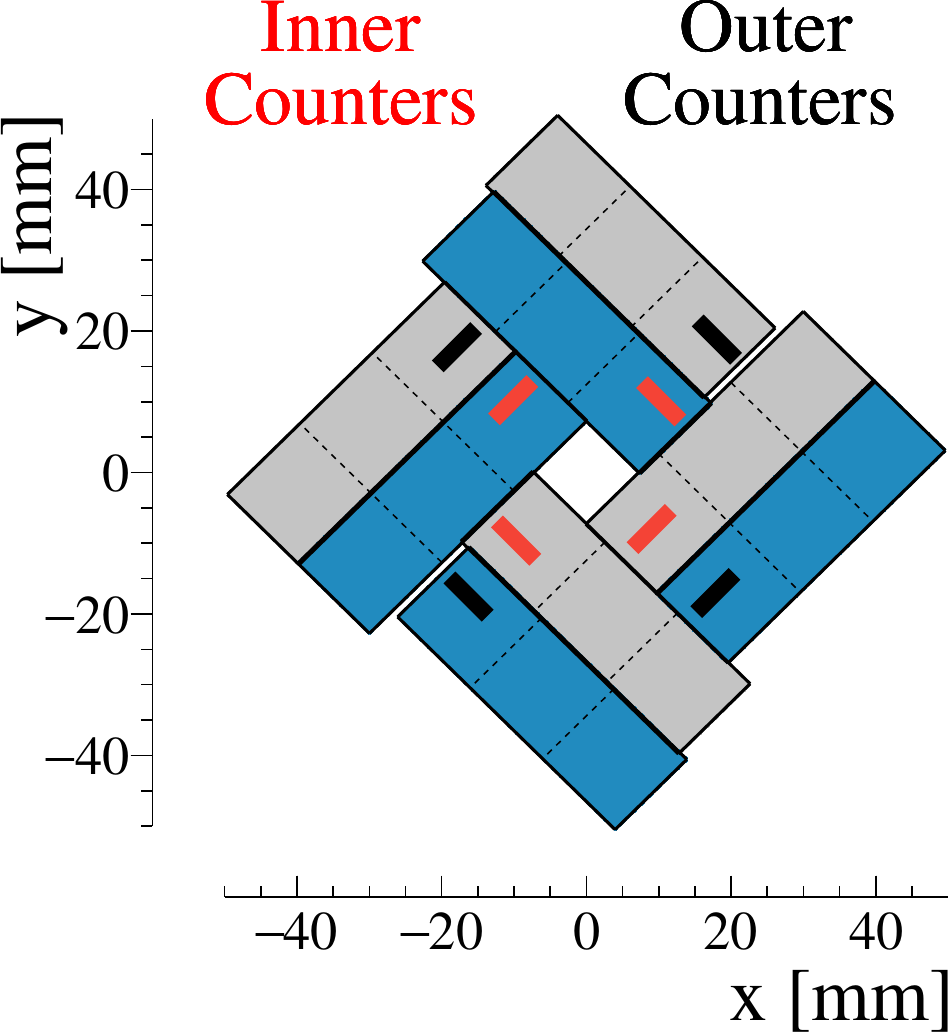}\hfill
\includegraphics[height=4.05cm]{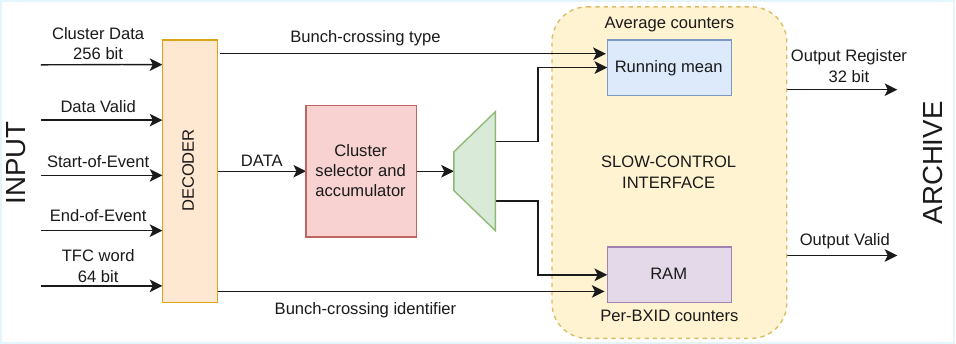}
\caption{(Left) Accumulation regions over the VELO sensors, where each square corresponds to one ASIC, and three ASICs read out one sensor. (Right) Schematics of the FPGA implementation of the luminosity accumulators. The input data are the reconstructed clusters for each event and the TFC word carrying the bunch-crossing type and identifier. The central block selects the clusters located within the accumulation regions, and performs the counting. The counts are then either averaged in a running mean, or simply stored, as explained in the text.}
\label{fig:counters}
\end{figure}

A set of cluster counters has therefore been added to the VELO TELL40 FPGA firmware~\cite{Hennessy:2021aec}. For each of the 26 layers of the VELO, four inner and four outer accumulation regions are defined, \ie one on each sensor, with the arrangement shown in Fig.~\ref{fig:counters}~(left). This amounts to a total of 208 counters. The inner counters lie at about 14\mm from the beam axis, while the outer ones at about 26\mm. Once calibrated, each counter provides a luminosity measurement that can be combined as needed to provide a single measurement. 
The accumulation regions chosen for this work have been verified to preserve linearity with respect to instantaneous luminosities up to five times the LHCb Run~3 working point~\cite{Passaro:2842603}.

At each event, the clusters recorded within the accumulation regions are counted. Four collision types can be identified, based on the LHC filling scheme and on whether both colliding bunches are filled with protons. If both are filled, it is a \textit{beam-beam} (bb); if neither, \textit{empty-empty} (ee). When only one bunch is filled with protons, the collision is either \textit{beam-empty} (be) or \textit{empty-beam} (eb), depending on which of the two beams has the filled bunch. 
Each pair of colliding bunches, whether filled or not, is classified with a bunch-crossing identifier (BXID) ranging from 0 to 3564.
Cluster counting is performed in two ways: separately per BXID, or averaged per collision type. The latter method uses a running mean to average over multiple events. The right-hand side panel of Fig.~\ref{fig:counters} shows an overview of the data flow within the relevant firmware blocks.

For each accumulation region $i$, the following types of accumulators are implemented in firmware:
\begin{itemize}
  \item \textit{average} counters $c_i$, \ie number of reconstructed clusters accumulated in the region $i$, separately for each bunch-crossing type: $c_i^{\rm bb}$, $c_i^{\rm eb}$, $c_i^{\rm be}$ and $c_i^{\rm ee}$;
  \item \textit{log0} counters $z_i$, \ie number of events where no clusters are found in the region $i$, separately for each bunch-crossing type: $z_i^{\rm bb}$, $z_i^{\rm eb}$, $z_i^{\rm be}$ and  $z_i^{\rm ee}$;
  \item \textit{per-BXID} counters $z_{ij}$, \ie number of events where no clusters are found in the region $i$, separately for each BXID $j$ among the 3564 composing the LHC orbit.
\end{itemize}
Among the accumulators presented in this paper, the average-method ones are chosen as main the luminosity estimators, and are combined into a single global estimator as discussed in Sect.~\ref{sec:results}.
Due to its robustness against saturation effects, the log0 method is used instead as a cross-check to assess the linearity of the average estimators.
Indeed, the log0 method is strongly based on the assumption that the number of the actual hits, whether detected or not,  follows a Poisson distribution. This is well verified when colliding bunches are taken individually. While saturation effects or hits inefficiencies do not affect the log0 method, any widening of the parent Poisson (for instance, due to varying expected number of hits) can introduce significant biases. This happens when the log0 counters are accumulated over several BXIDs. In this case, the number of interactions is no longer distributed according to a Poisson function, but rather as the superposition of several independent Poisson distributions with different means, for which Eq.~\ref{eqn:log0} no longer holds. For this reason, the average counters are preferred.

The per-BXID accumulators use the log0 method because of the limited availability of FPGA resources. The implementation of per-BXID accumulators in the firmware requires the usage of RAM blocks to store the value of each BXID \muvis.
In order to avoid excessive memory resource usage, the RAM words are 20~bits long. If these counters used the average method, it would be technically possible for them to go out of range even when integrating less than $2^{20} \approx 10^6$ events. Using the log0 method, the accumulator value is equal at most to the number of integrated events, preventing the accumulator from overflowing.
Also due to the limited availability of FPGA resources, the log0 and per-BXID accumulators have been implemented only for the 104 outer accumulation regions.

For the average and log0 accumulators, a running mean $\hat{x}$ is computed by counting hits over a fixed number of events $N$ at each step $t$, then averaging the most recent counter value $x$ with the previous value of the running mean, weighted by factors $\lambda$ and $1-\lambda$, respectively, as\footnote{The running-mean method is not strictly needed for providing a luminosity measurement. It has been implemented in order to be as independent as possible from the readout architecture, while at the same time providing accumulators that are as memory-efficient as possible and with the best resolution.}
\begin{equation}
    \hat{x}_t = \lambda\,\hat{x}_{t-1}+ (1-\lambda)\, x_t.
\end{equation}
\begin{figure}
    \centering
    \raisebox{-.5\height}{\includegraphics[width=0.49\linewidth]{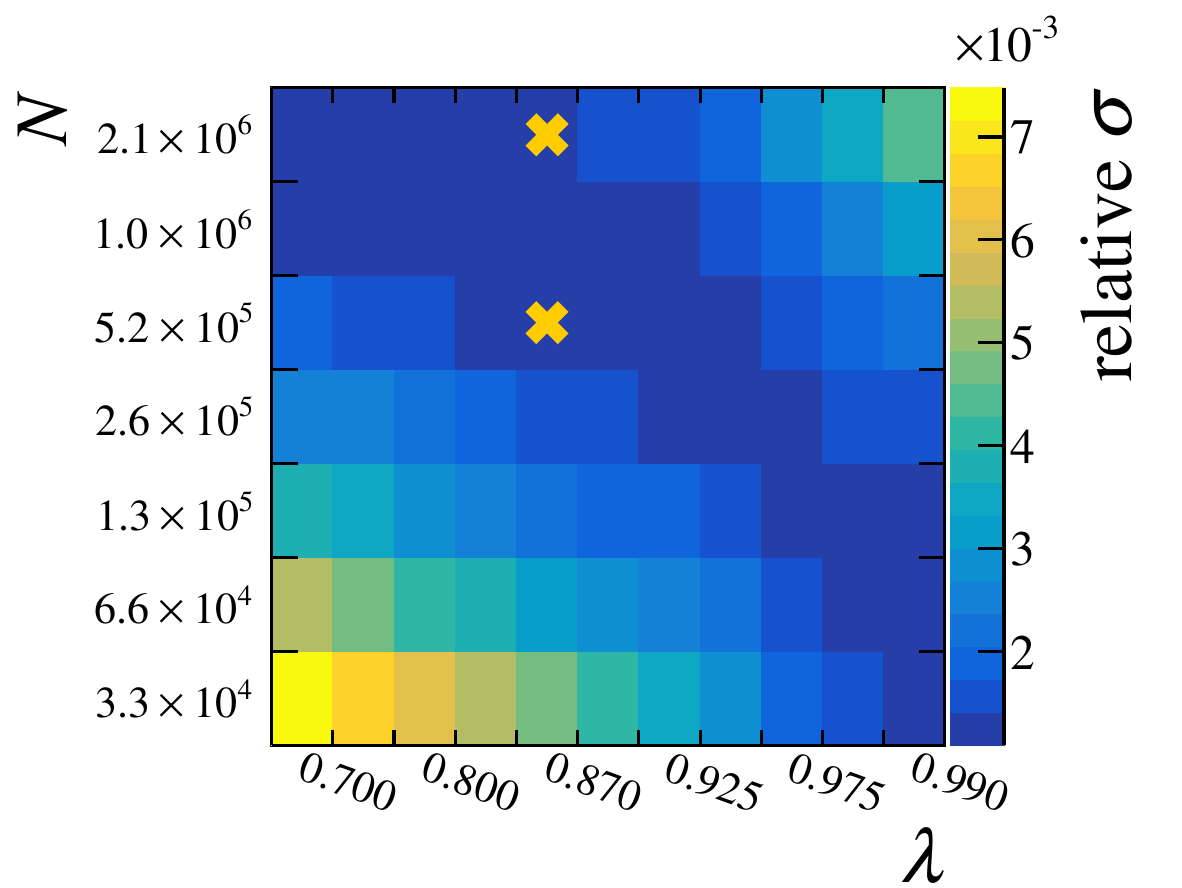}}
    \raisebox{-.5\height}{\includegraphics[width=0.49\linewidth]{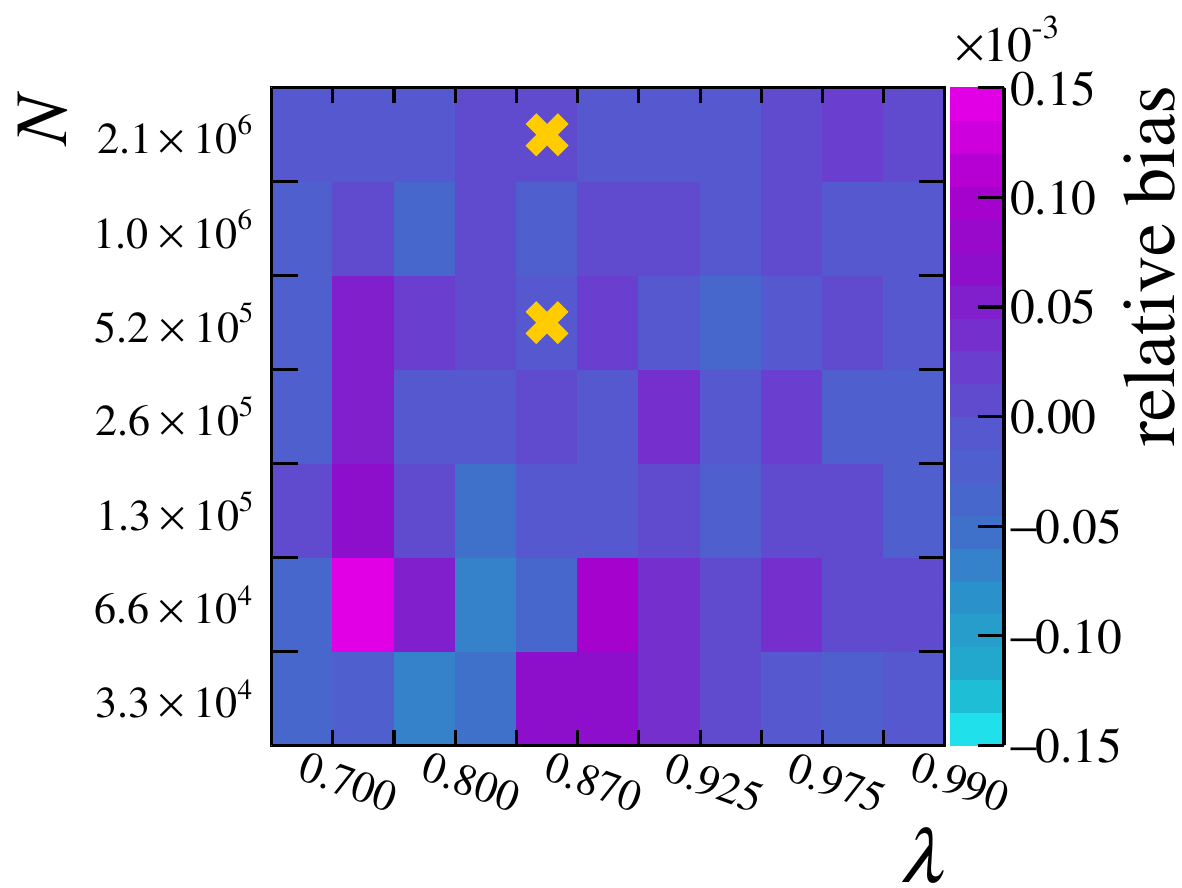}}\\[7pt]
    \raisebox{-.5\height}{\includegraphics[width=0.45\linewidth]{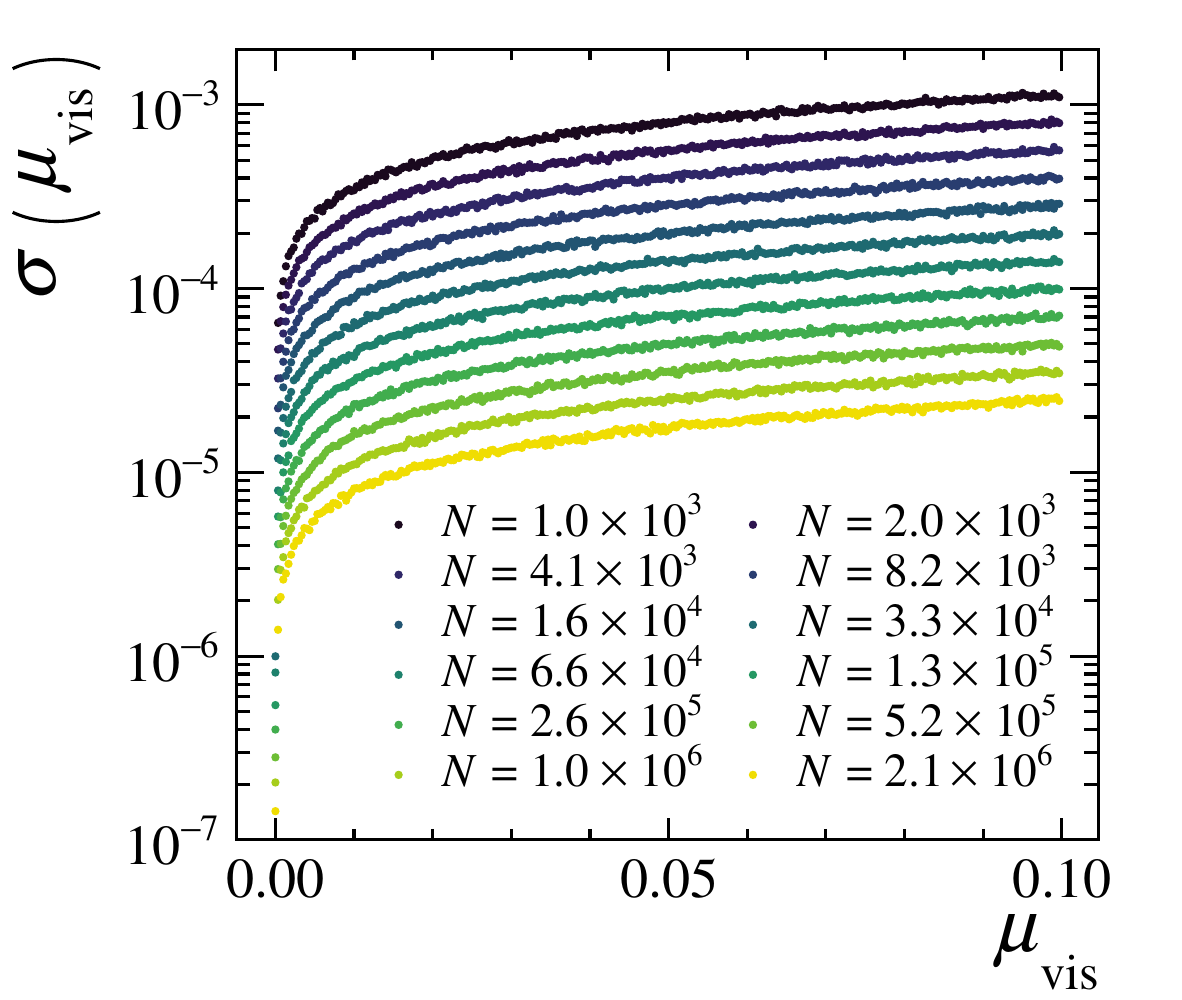}}
    \caption{(Top left) Relative standard deviation and (top right) relative bias of a running-mean estimator as a function of the $N$ and $\lambda$ parameters, for a typical occupancy of $\muvis=0.1$ representative of an inner accumulation region at nominal Run~3 luminosity. The yellow markers represent the parameters used for \proton\proton (upper) and PbPb (lower marker) data taking. (Bottom) Absolute standard deviation of the running-mean estimator as a function of \muvis, for different $N$ values and fixed $\lambda$. The results shown in all three panels are obtained using simulated pseudoexperiments assuming a scenario of levelled luminosity at the Run~3 working point. Similar results are obtained assuming a scenario where the luminosity is let to exponentially decay.}
    \label{fig:running-mean}
\end{figure}

Pseudoexperiments are used in order to confirm the absence of bias or degraded resolution in different running conditions, testing both exponentially decaying and levelled luminosity. In these pseudoexperiments, which take into account the expected occupancy on the VELO sensors at a given \proton\proton luminosity, the hit counts are accumulated at the same frequency of the hardware implementation, and the running mean is then calculated for varying values of the parameters $N$ and $\lambda$. 
The bias and resolution of the running-mean estimator are computed as the mean value and standard deviation of the distribution of the relative difference between reconstructed and simulated luminosity.
The first two panels of Fig.~\ref{fig:running-mean} show the resolution and bias on \muvis for varying running-mean coefficients.
The integration parameters used throughout the 2024 data-taking period have been $\lambda = 0.875$ and $N=1024 \times N_{\rm bb}$, \ie $N$ is dynamically set based on the LHC filling scheme. This corresponds to updating the running mean every $\sim$90\ms, \ie to counting hits over $N\simeq 2.18\times 10^6$ bunch-crossing events in nominal conditions for \proton\proton physics data taking in Run~3.
The running-mean-based accumulators are read out every 3\sec via ECS, while the per-BXID ones require a longer integration window of about 40\sec to accumulate enough data, and are then read out every 3~minutes. 
The values of both types of accumulators are stored in a database at every reading and can be used for later analysis.
The reading frequency is driven by the choice of performing real-time analysis in the ECS software. It could be increased if counter data were streamed through a different data path and analysed using additional dedicated resources.

The quantities used as luminosity counters are finally calculated in ECS, based on the value of the firmware accumulators, performing a background subtraction (the ee contribution is added to avoid double counting~\cite{LHCb-PAPER-2014-047}): 
\begin{itemize}
    \item average method: $\mu_{\text{vis}, i}^\text{avg} = \dfrac{c_i^{\rm bb}}{N_{\rm bb}} -\dfrac{c_i^{\rm be}}{N_{\rm be}} - \dfrac{c_i^{\rm eb}}{N_{\rm eb}} + \dfrac{c_i^{\rm ee}}{N_{\rm ee}}$, where the index $i$ runs across the 208 accumulators and $N_{k}$ is the number of collected events for each bunch-crossing type $k$;
    \item log0 method: $\mu_{\text{vis}, i}^\text{log0} = -\left(\ln\dfrac{z_i^{\rm bb}}{N_{\rm bb}} -\ln\dfrac{z_i^{\rm be}}{N_{\rm be}} - \ln\dfrac{z_i^{\rm eb}}{N_{\rm eb}} + \ln\dfrac{z_i^{\rm ee}}{N_{\rm ee}}\right)$;
    \item per-BXID: $\mu_{\text{vis},ij} = -\ln$ $\dfrac{z_{ij}}{M_j}$, where the index $j$ runs across the 3564 BXIDs and $M_j$ is the total number of events with BXID $j$.
\end{itemize}

Further pseudoexperiments are used to assess the statistical uncertainty of the running-mean-based estimators. The results are shown in the bottom panel of Fig.~\ref{fig:running-mean}. The standard deviation $\sigma\left(\muvis\right)$ on the luminosity counter $\muvis$ is found to be empirically well described by a square-root law, with $\sigma\left(\muvis\right)\approx (1-\lambda)\sqrt{\muvis/N}$.

The luminosity counters presented in this paper are implemented in firmware and software, and come thus with a flexibility not normally available for readout-level counters based on currents or other detector quantities. By tuning their position, shape, size and integration parameters, the accumulation regions can be optimised to fit occupancy requirements, bandwidth, and constraints due to the usage of FPGA resources. In particular, the counters can be tuned to handle a wide dynamic range, from very low to very high luminosities. This currently requires recompiling the VELO DAQ firmware; however, dedicated registers can also be implemented, to dynamically set the counter shapes and integration windows. The overall FPGA resource usage for the current implementation amounts to 0.62\% of the $1.15 \times 10^6$ logic elements, 2.4\% of the 2713 M20K memory blocks and 2.8\% of the 1518 digital signal processors (DSP). The M20K memory blocks are used for the per-BXID measurement, and the DSPs for the running-mean computation.

\section{Calibration}
\label{sec:calib}
The instantaneous luminosity is calculated from \muvis and \svis as shown in Eq.~\ref{eqn:lumi}.
Each of the time-integrated luminosity counters is independently calibrated by measuring the associated visible cross-section \svis; each counter thus in principle provides an independent luminosity measurement.\footnote{The counters are slightly correlated with each other, due to their calibrations all relying on the same measurement of the beam current as input.}
Two calibration techniques are used at LHCb. One of these, used in Run~1 and Run~2 and also capable of assessing the rate of beam background~\cite{Coombs:2021}, is the beam-gas imaging technique~\cite{FerroLuzzi:2005em,Barschel:2014iua}, based on the measurement of the overlap integral of the two beams from their shapes. The other one, historically used at hadron colliders, relies on dedicated data-taking periods called van der Meer (vdM) scans~\cite{vanderMeer:296752,Balagura:2020fuo}. The work presented in this paper uses the latter method.

During a vdM scan, \svis is measured by displacing the two beams with respect to one another in the plane transverse to the beam direction, recording the rate of visible interactions at each $\left(\Delta x, \Delta y\right)$  step, and integrating its two-dimensional profile as
\begin{align}
    \svis = \iint\mupvis\left(\Delta x,\Delta y\right) \mathrm{d}\Delta x \, \mathrm{d}\Delta y \, ,
    \label{eqn:vdm2d}
\end{align}
where $\mupvis\equiv\muvis/n_1n_2$, and $n_{1,2}$ are the populations of the colliding bunches measured at each step. The per-bunch intensities are inferred using data from the Fast Bunch Current Transformers (FBCT)~\cite{Belohrad:1267400,Anders:1427726} normalised with the beam-current measurement provided by the LHC Direct Current Current Transformers (DCCT)~\cite{Barschel:1425904}.
Under the assumption that the overlap integral follows a Gaussian distribution, Eq.~\ref{eqn:vdm2d} reduces to
\begin{align}
    \svis = 2\pi\mupvis(0,0)\Sigma_x\Sigma_y , \
    \label{eqn:vdm2d_gaus}
\end{align}
where $\mupvis(0,0)$ is the population-normalised \muvis value when the beams are not displaced, and $\Sigma_{x}$ and $\Sigma_{y}$ are the standard deviations of the scan profile on the $x$ and $y$ axes, respectively, \ie the convolved projected widths of the two beams~\cite{Balagura:2020fuo, Babaev:2023fim}.

The visible cross-section is specific to each luminosity counter, and its value and physical interpretation depend on the counter properties.
In the case of the log0 method, $\svis^{\text{log0}}$ is related to the inelastic \textit{pp} cross-section by $\sigma_{\text{vis}}^{\text{log0}} = \varepsilon\sigma_{pp}^{\text{inel}}$, where $\varepsilon$ encompasses the geometrical acceptance and efficiency of the active sensor surface used as the luminosity counter. Therefore, $\svis^{\text{log0}}$ can be interpreted as a physical cross-section. 
Conversely, in the case of the average method, \mbox{$\svis^{\text{avg}} = k\varepsilon\sigma_{pp}^{\text{inel}}$}, where the factor $k$ is related to the properties of the physical quantity used as the luminosity counter (in this case, the average number of clusters per inelastic \proton\proton collision).
Therefore, $\svis^{\text{avg}}$ cannot be directly interpreted as the cross-section of a physical process, but rather as a proportionality factor between visible interactions and instantaneous luminosity.
As a consequence, $\svis^{\text{avg}}$ and $\svis^{\text{log0}}$ are not expected to have the same value, even for the same accumulation region.

\begin{figure}
    \centering
    \includegraphics[width=\linewidth]{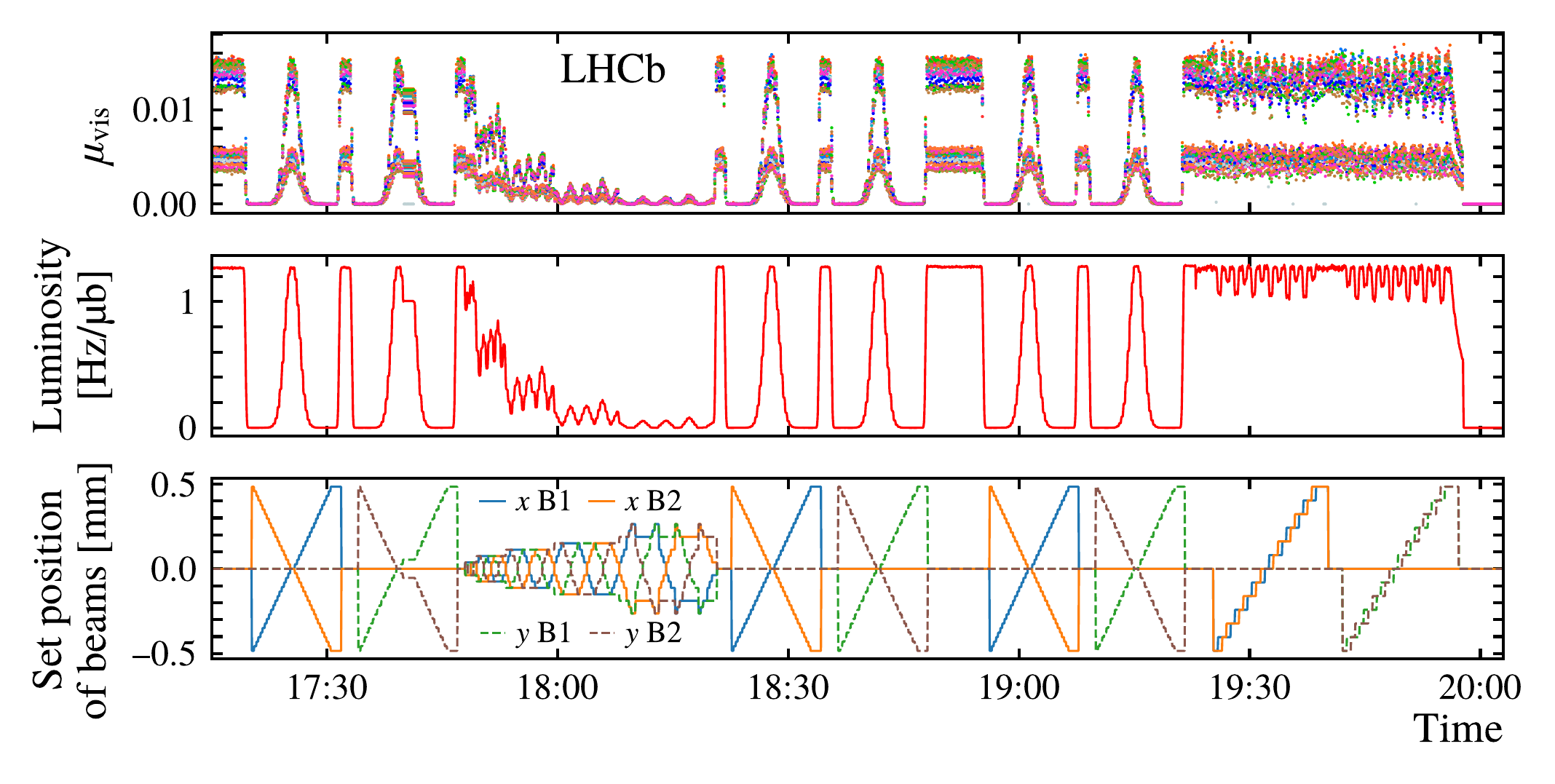}
    \caption{Snapshot of the \proton\proton 2024 vdM scan. (Top) Number of visible interactions $\muvis$ as computed from a randomly chosen subset amounting to 1/3 of all available counters (for display reasons). Each colour represents a different counter. The rate difference between the inner and outer accumulation regions is clearly visible. (Center) Global luminosity estimator, obtained via a trimmed mean of all luminosity measurements (see Sect.~\ref{sec:results} and Fig.~\ref{fig:trim}). The tiny luminosity drops visible during the LSC phase, \ie the last $\sim$40 minutes, are due to the asynchronous movements of the beams, thus to short periods of time during which the beams were not perfectly head-to-head. (Bottom) Positions of the two beams, as set by LHC control, as a function of time throughout the scan.}
    \label{fig:vdm_all_counters}
\end{figure}

Figure~\ref{fig:vdm_all_counters} illustrates the various phases of the 2024 \proton\proton vdM scan. 
The bottom panel shows the positions of the two beams along $x$ and $y$ during a typical sequence.\footnote{A standard vdM sequence in LHCb consists of a first set of one-dimensional scans performed separately along $x$ and $y$, followed by a two-dimensional scan in which the beams are simultaneously shifted along both directions, and by another set of one-dimensional (1D) scans. A third 1D scan is performed, usually with argon injected in the SMOG2 storage cell, followed by a length-scale calibration (LSC)  during which the beams are shifted head-to-head in order to calibrate the size of the vdM displacement steps. In the particular sequence shown in Fig.~\ref{fig:vdm_all_counters},  gas injection was performed between the third 1D scan and the LSC scan.} The top panel shows the interaction rate measured by a subset of all available counters. 

A full vdM sequence, which requires dedicated beam optics, lasts several hours and it is only performed once per year, per centre-of-mass energy and beam type~\cite{LHCb-PAPER-2014-047}. However, luminometer calibrations can be checked and updated during similar, much shorter \textit{emittance scans}, which can be performed in a few minutes at the end of an LHC fill using the same beam optics as in physics data taking~\cite{Emittance.PhysRevAccelBeams}. These scans have fewer steps than a full vdM sequence, due to time constraints, and have a much higher \muvis, which may not be safe against saturation effects.

\begin{figure}
    \centering
    \includegraphics[width=.57\linewidth]{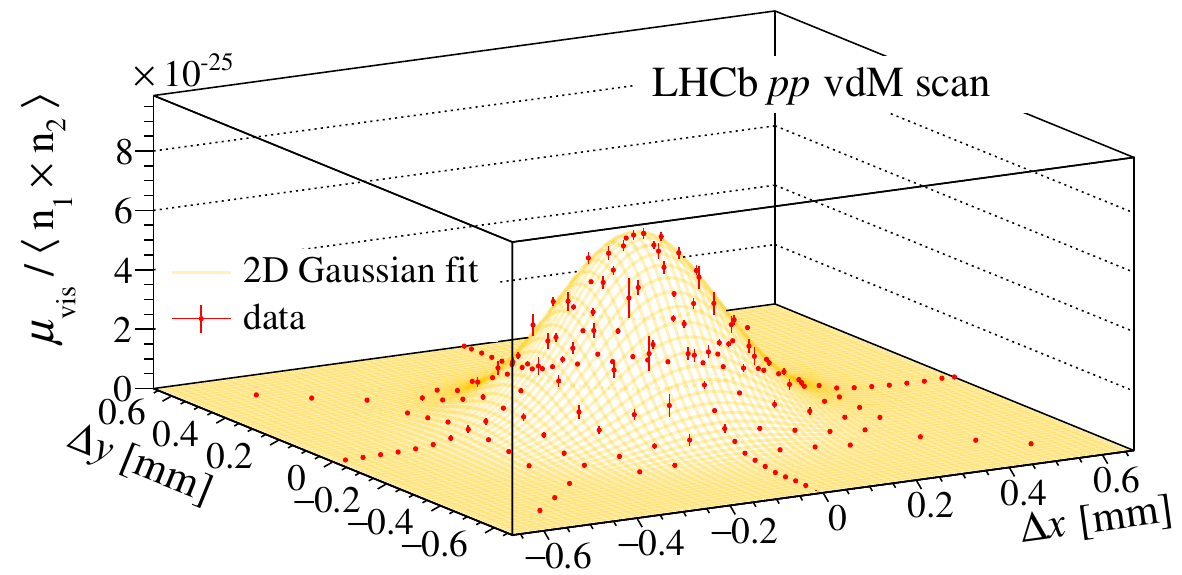}\hfill
    \includegraphics[width=.42\linewidth]{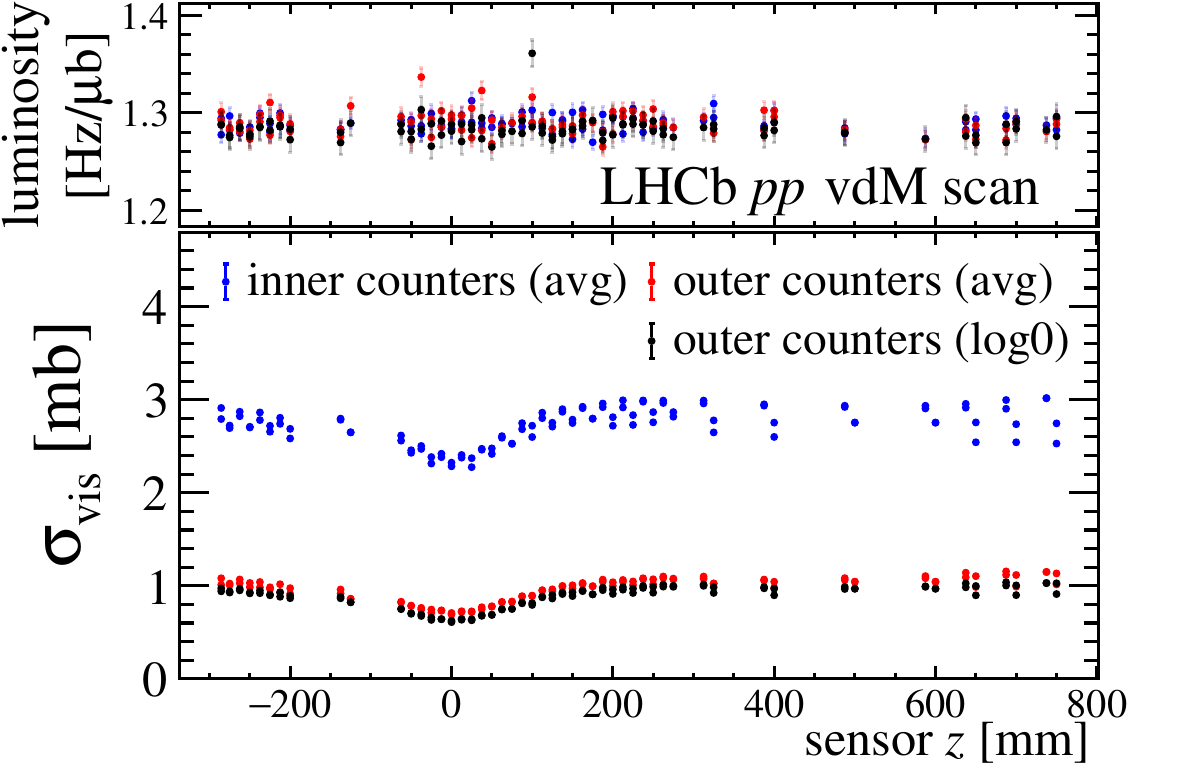}
    \caption{(Left) Example of \proton\proton vdM calibration result for an \textit{average}-method counter from one of the outer accumulation regions.
    The data are background subtracted, and normalised with the average populations of the colliding bunches. Uncertainties are statistical only. 
    (Right) Each point represents the visible cross-section measured by one of the counters. The three data series correspond to different methods and to either inner or outer counters. The ``valley-like'' shape reflects the geometrical acceptance of the different layers of the VELO detector. The top subpanel gathers the calibrated luminosity measurements obtained from all of the available counters during a period of quasi-constant luminosity during the vdM programme, with the same colour scheme.}
    \label{fig:vdm24pp}
\end{figure}

\begin{figure}
    \centering
    \includegraphics[width=.57\linewidth]{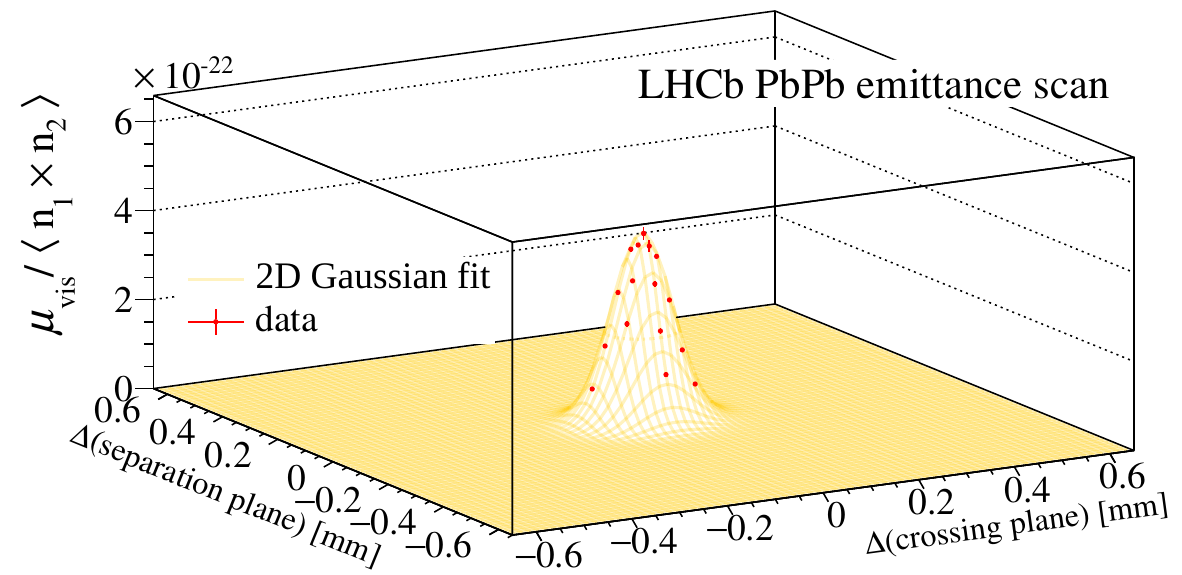}\hfill
    \includegraphics[width=.42\linewidth]{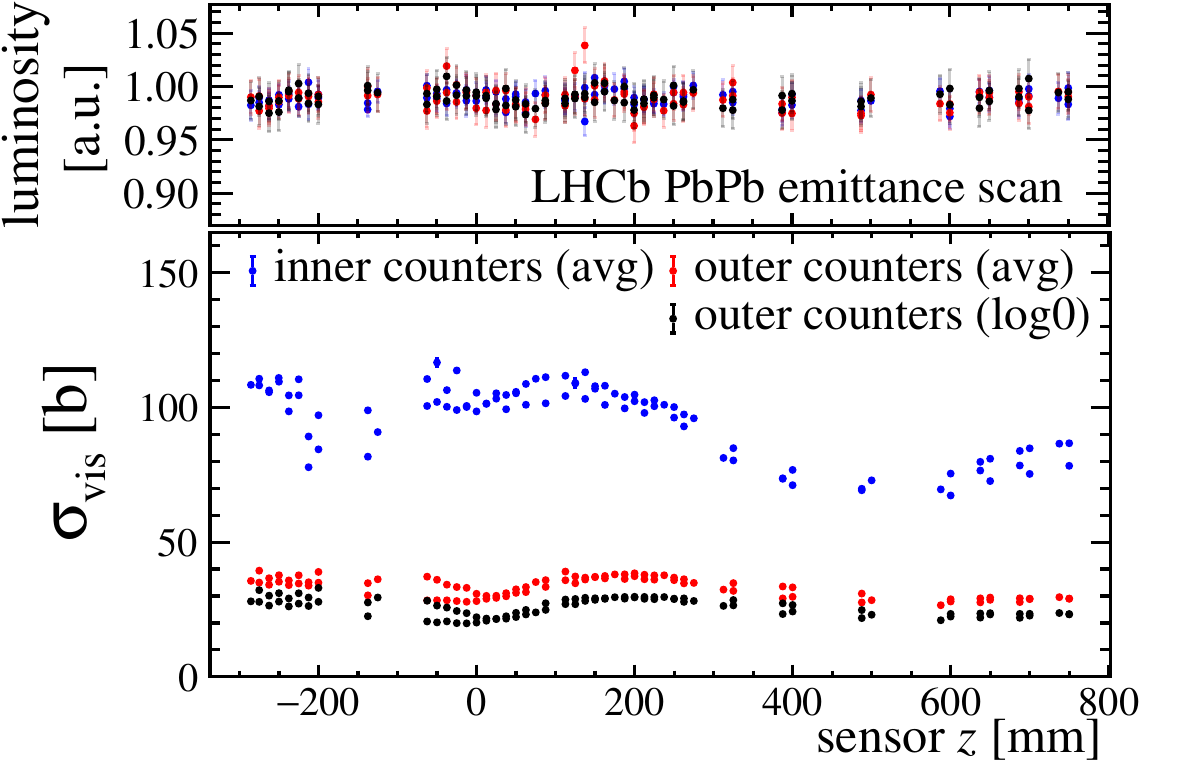}
    \caption{Same as Fig.~\ref{fig:vdm24pp}, but performed with PbPb collisions data collected during a dedicated emittance scan. The \muvis profile shown on the left is narrower than the profile observed in Fig.~\ref{fig:vdm24pp}, because of the different beam optics used during the emittance scan in PbPb collisions. On the right-hand side panel, the luminosity units shown are arbitrary, as a quantitative study of the PbPb luminosity requires further work that exceeds the scope of this paper. 
    }
    \label{fig:vdm24leadlead}
\end{figure}

Figure~\ref{fig:vdm24pp} (left) shows the result of the calibration performed using data from the 2024 vdM scan with \proton\proton collisions at centre-of-mass energy $\sqs=13.6\tev$.
The distribution of counter rates is modelled with a 2D Gaussian probability density function. The cross-section is obtained according to Eq.~\ref{eqn:vdm2d_gaus}. 
Figure~\ref{fig:vdm24pp} (right) gathers the computed visible cross-sections for all the luminosity counters introduced in Sect.~\ref{sec:counters}, 
arranged by the $z$ coordinate of the VELO module on which the selection region is located. The trend visible in the figure reflects the geometrical acceptance of the VELO~\cite{Passaro:2842603}. This panel also gathers the luminosity values obtained after the calibration from all of the available counters. Figure~\ref{fig:vdm24leadlead} shows a similar calibration for PbPb collisions in 2024. The data used here were collected during two one-dimensional emittance scans, combined together for the analysis.
The $z$-dependence of the calculated cross-sections is significantly different with respect to the \proton\proton case. The variation in the acceptance of the VELO modules can be attributed to the electromagnetic processes that dominate the PbPb cross-section, highly enhanced with respect to \proton\proton collisions~\cite{Bertulani_2005}. 

For \proton\proton collisions, the calibrated luminosity measurements obtained during the vdM scan, and shown on top of the right-hand side panel of Fig.~\ref{fig:vdm24pp}, are in good agreement with each other, showing a relative spread of $0.8\%$, including inner, outer and log0 counters.
Similarly, in the PbPb case, the calibrated luminosities show a relative spread of $0.9\%$.
Systematic uncertainties associated with the movements of the interaction region along the $z$ axis, coupled with the beam displacement on the $xz$ plane, are measured to be $\mathcal{O}(0.1\%)$ during the calibration procedure. This contribution is thus considered negligible. Further discussion on the systematic uncertainties can be found in Sect.~\ref{sec:systs} as well as in Refs.~\cite{LHCb-PAPER-2014-047, Balagura:2020fuo, CMS:2021xjt}.

\section{Results}
\label{sec:results}

\begin{figure}
    \centering
    \hspace{-3mm}\includegraphics[height=0.37\linewidth]{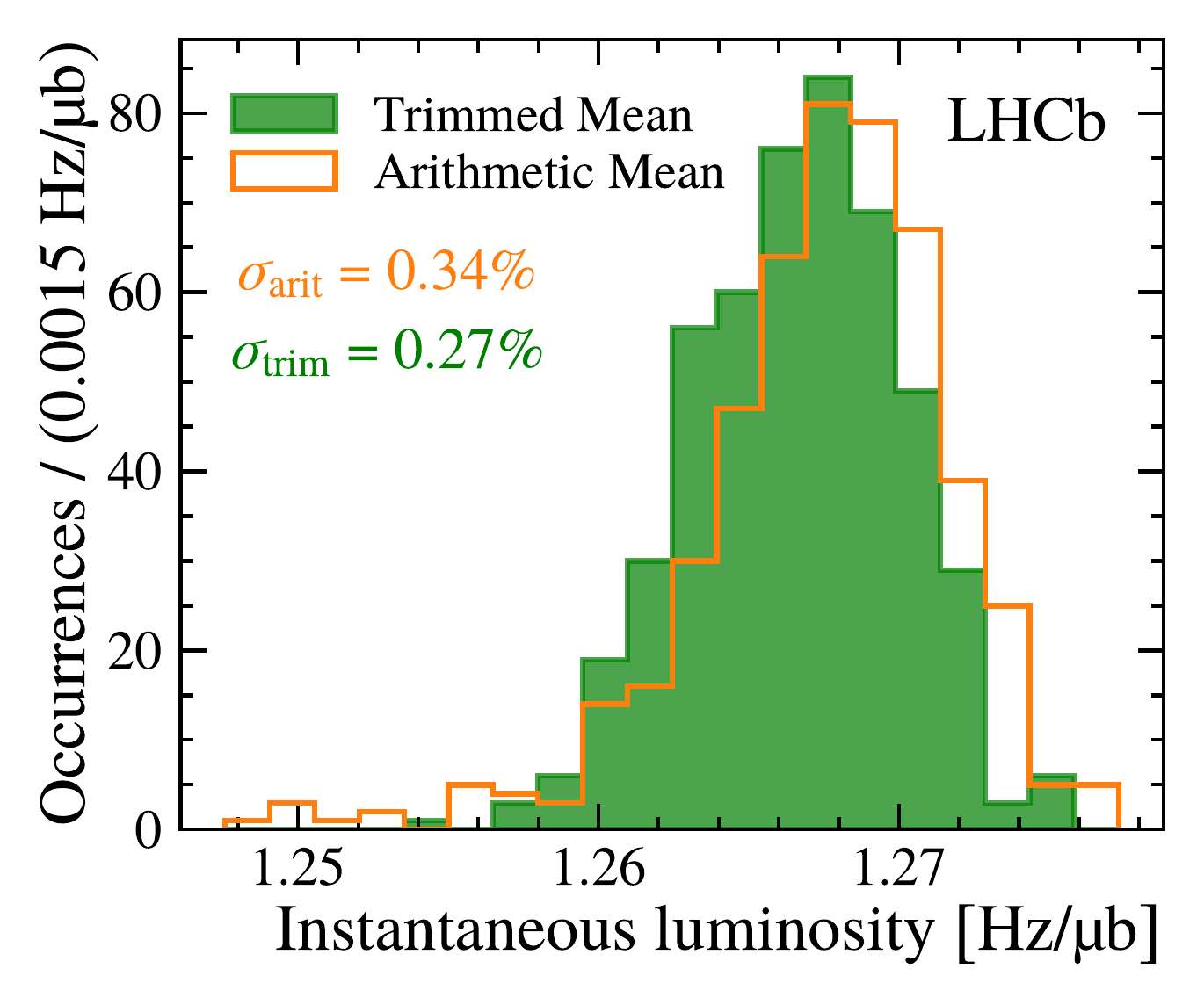}\hspace{3mm}
    \includegraphics[height=0.37\linewidth]{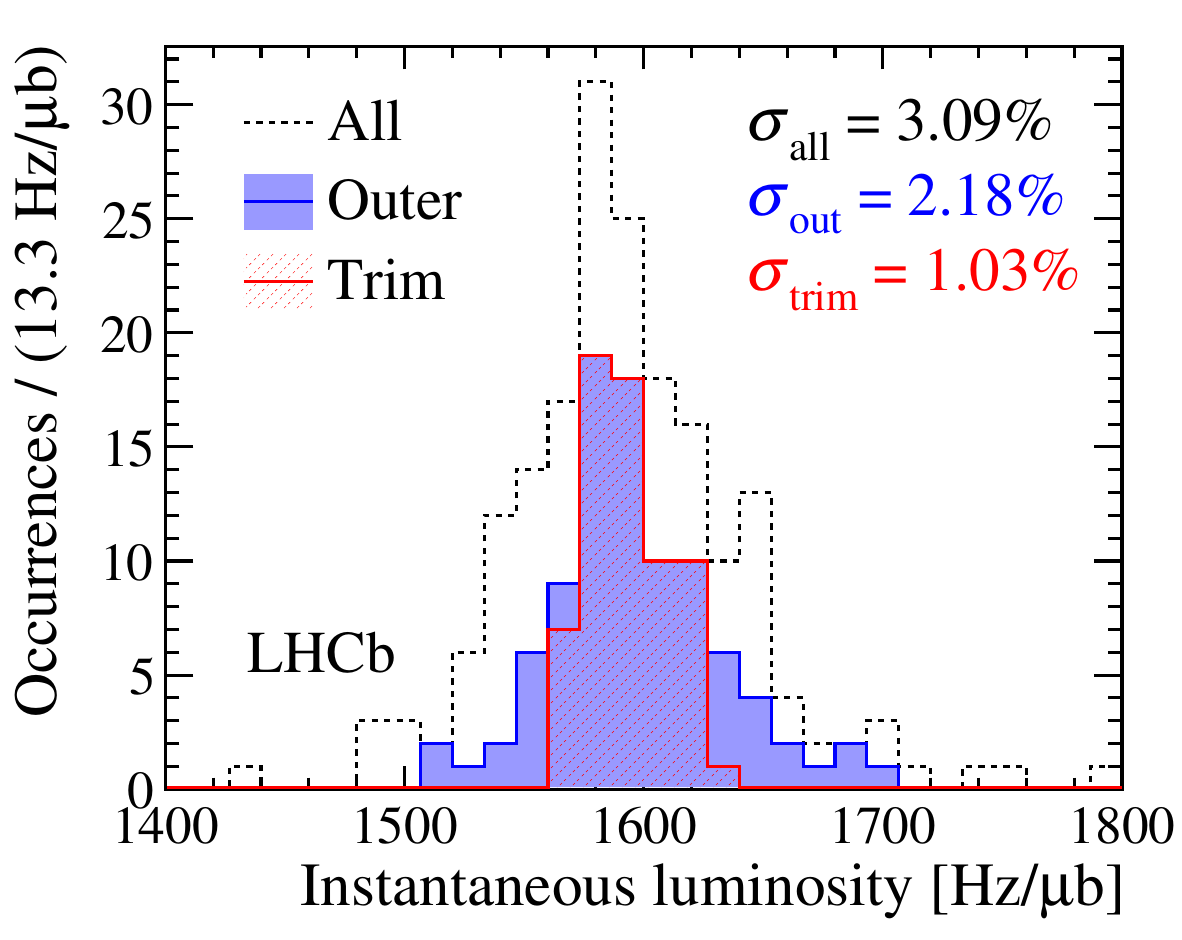}

    \caption{(Left) Distributions of the luminosity estimators on \proton\proton collisions gathered over a 25 minute-long period of approximately constant luminosity during the 2024 vdM scan. Each entry corresponds to a different luminosity value, measured every 3\sec. In green, the trimmed mean of the counters used as global luminosity estimator. Overlaid in orange, the distribution of the arithmetic mean of the counters over the same time period. (Right) Distribution of the luminosity estimates in a 30\sec time interval during physics data taking, from \proton\proton collisions recorded in July 2024. The dotted histogram gathers data from all available counters, while the solid blue histogram contains only the outer counters. Overlaid, the red striped histogram shows the resulting distribution once the threshold for the trimmed mean is applied to the set of outer counters. }
    \label{fig:trim}
\end{figure}

Figures~\ref{fig:vdm24pp} (right) and~\ref{fig:vdm24leadlead} (right) show that the luminosity measurements obtained using the various counters, even from different VELO sensors, are consistent with each other within about 1\%.
The 208 counters can be combined into a single and more robust global luminosity estimator.
Three statistical combinations are tested: mean, median and a trimmed mean where the 15\% most extreme values on each side are removed. The performance of these estimators is studied using a sample of data from the 2024 \proton\proton vdM programme with an almost constant luminosity. The trimmed mean estimator, shown in Fig.~\ref{fig:trim} (left), is chosen as the main luminometer because of its intrinsic robustness to outliers. 
The trimmed mean estimator has a mean value within 0.1\% of the arithmetic mean, but a 20\% smaller relative standard deviation, and a more symmetrical shape with significantly reduced tails.

Only the outer average counters are inserted in the trimmed mean.
During data taking for physics, the beam conditions at the interaction point differ from the vdM case in terms of number of colliding bunches, beam parameters, including \eg bunch shape and crossing angle, and stability of the luminous region position~\cite{LHCb-PAPER-2014-047}. 
As the sensitivity to such variations decreases as the distance from the luminous region increases, a suitable choice for the global estimator is to only insert the outer counters in the trimmed-mean estimator. Other possible choices could be to pick the counters located on the most upstream and downstream VELO stations, or to construct a distance-weighted estimator. For the sake of simplicity, hereinafter the performance of the first option is analysed. To corroborate this choice, Fig.~\ref{fig:trim} (right) shows the distribution of the luminosities measured in physics data-taking conditions by all counters, by the outer counters only, and by those outer counters that are dynamically selected within the trimmed mean estimator. This comparison uses an example dataset collected during a short period of time, in which the beam conditions can be assumed to be constant. As expected, the relative spread of the measurements shrinks when using the outer counters only. Therefore, this strategy is adopted throughout the rest of this paper.

The global luminosity estimator has been used to measure the instantaneous luminosity at LHCb throughout the 2024 data-taking period. In the following sections, the performance of this estimator is discussed and, where relevant, compared to that of the LHCb luminometer PLUME. The statistical resolution, linearity, and stability in time are presented. The response to gas injection in SMOG2 is discussed, as well as the consistency of the luminosity counters with respect to each other and to PLUME. The viability of the method in PbPb collisions is assessed. Finally, possible systematic effects are introduced and discussed, as well as the remarkably small downtime.

\subsection{Statistical resolution}
An upper limit on the counter resolution is evaluated by acquiring several measurements at --in principle-- the same luminosity.\footnote{At a particle collider, truly constant luminosity cannot be achieved due to beam degradation: the population of colliding bunches strictly decreases with time. Luminosity levelling mitigates this effect by acting on the beam optics, but it is by design not a strictly continuous procedure.} Figure~\ref{fig:trim} (left) shows the luminosities measured by the trimmed mean estimator over a period of 25~minutes of approximately constant luminosity during the 2024 \proton\proton vdM programme. The relative spread of these measurements, $\sigma_\mu/\mu=0.27\%$, can be interpreted as an upper bound on the statistical resolution of the global luminosity estimator. This value is partly due to the intrinsic statistical resolution and partly to fluctuations of the luminosity itself. 
Furthermore, it should be noted that this value has been evaluated at low luminosity during a vdM programme, with an average occupancy at the level of $10^{-2}$ clusters per event. In normal physics data-taking conditions, the higher number of colliding bunches means that the counters integrate more  events within the same time window. This is about 100 times more events, considering the filling schemes used during the 2024 data-taking run. This leads to better statistical resolution. The scaling of the latter with respect to $\mu$, and therefore to the luminosity, has been shown for a single counter in Fig.~\ref{fig:running-mean} (bottom).
With the assumption that the correlation between the counters does not depend on luminosity, one can expect the same scaling for the global estimator, \ie a $\mathcal{O}(10^{-4})$ statistical resolution.

\subsection{Linearity}

\begin{figure}
    \centering
    \includegraphics[width=0.328\linewidth]{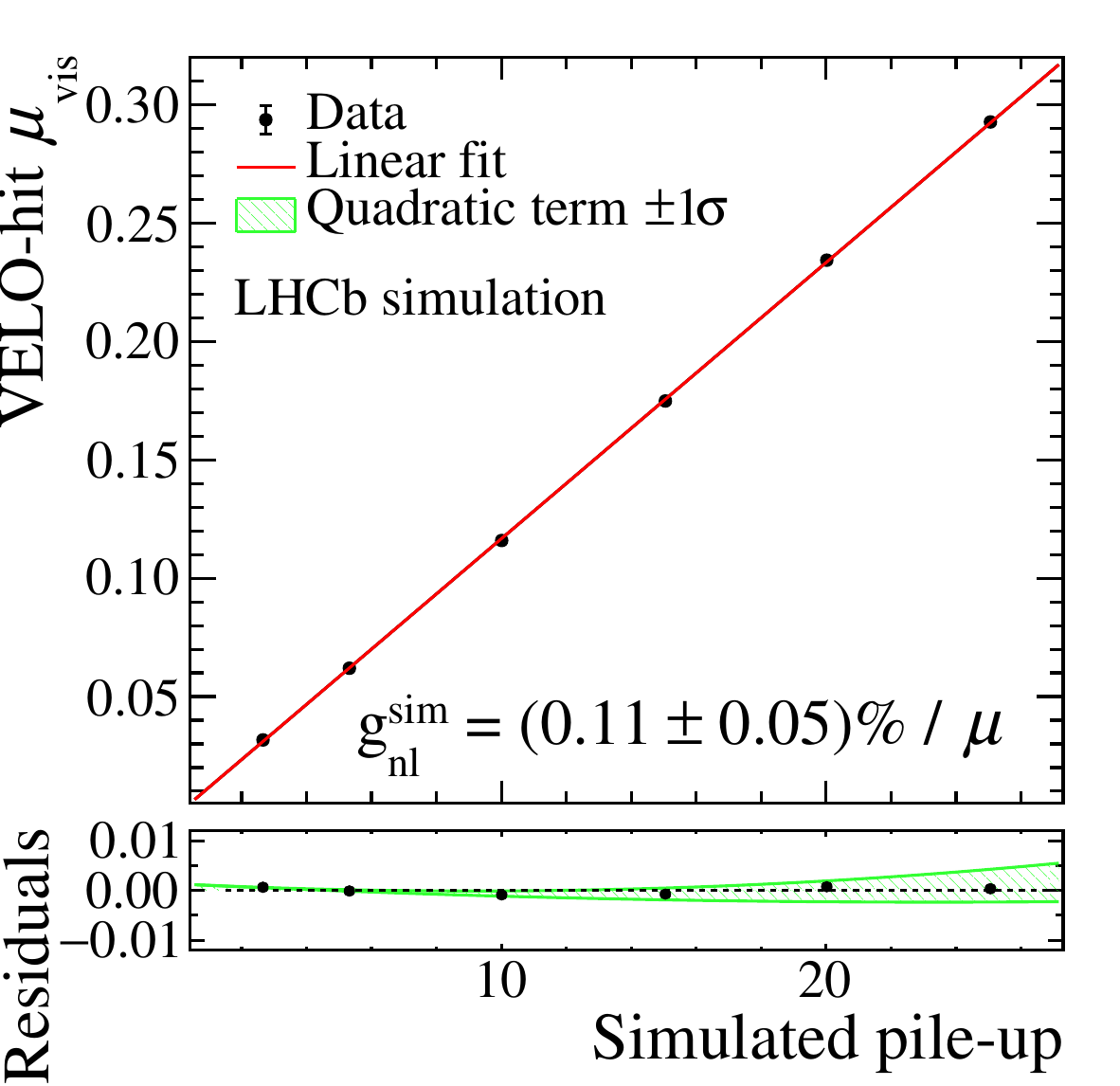}
    \includegraphics[width=0.328\linewidth]{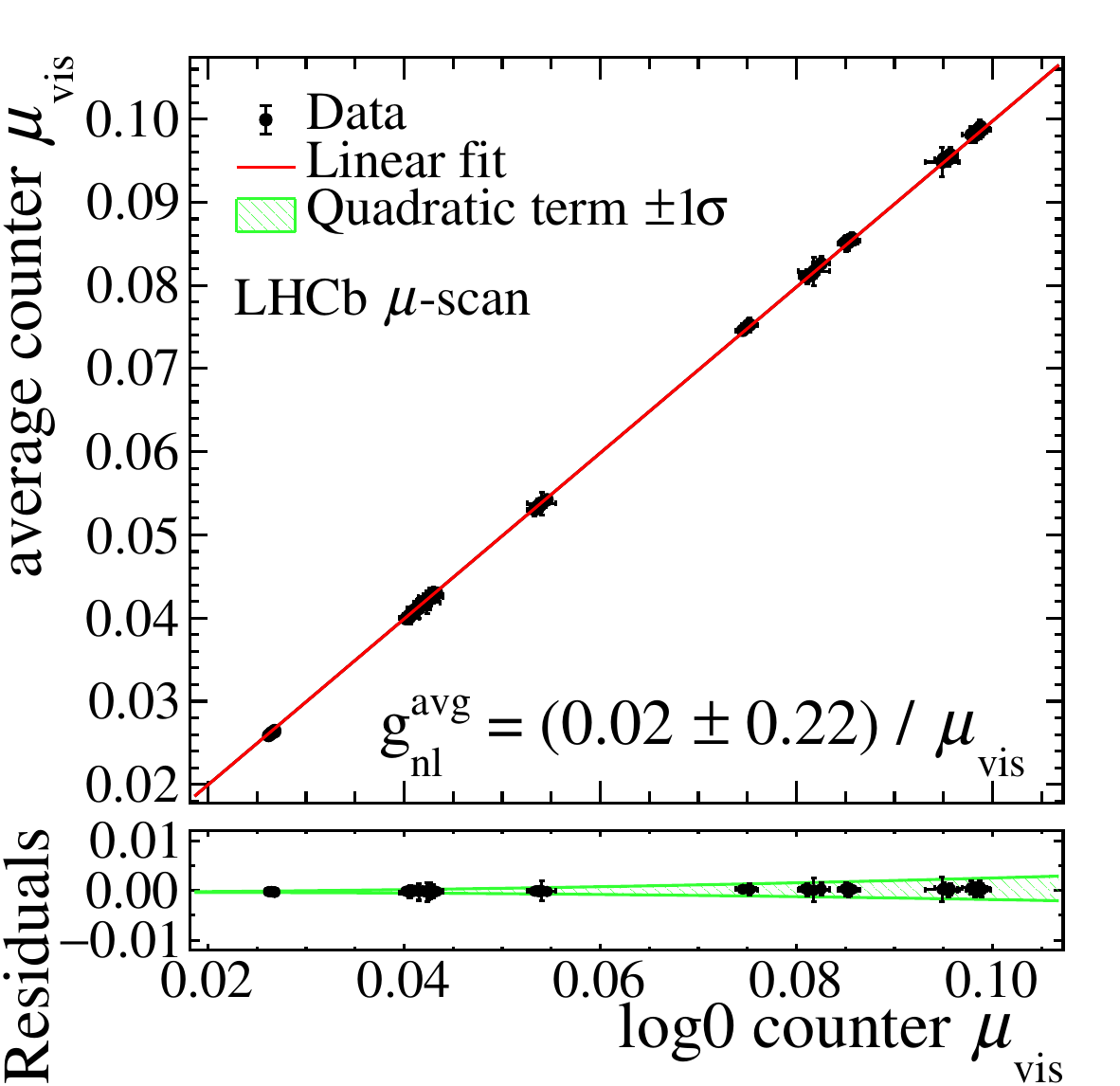}
    \includegraphics[width=0.328\linewidth]{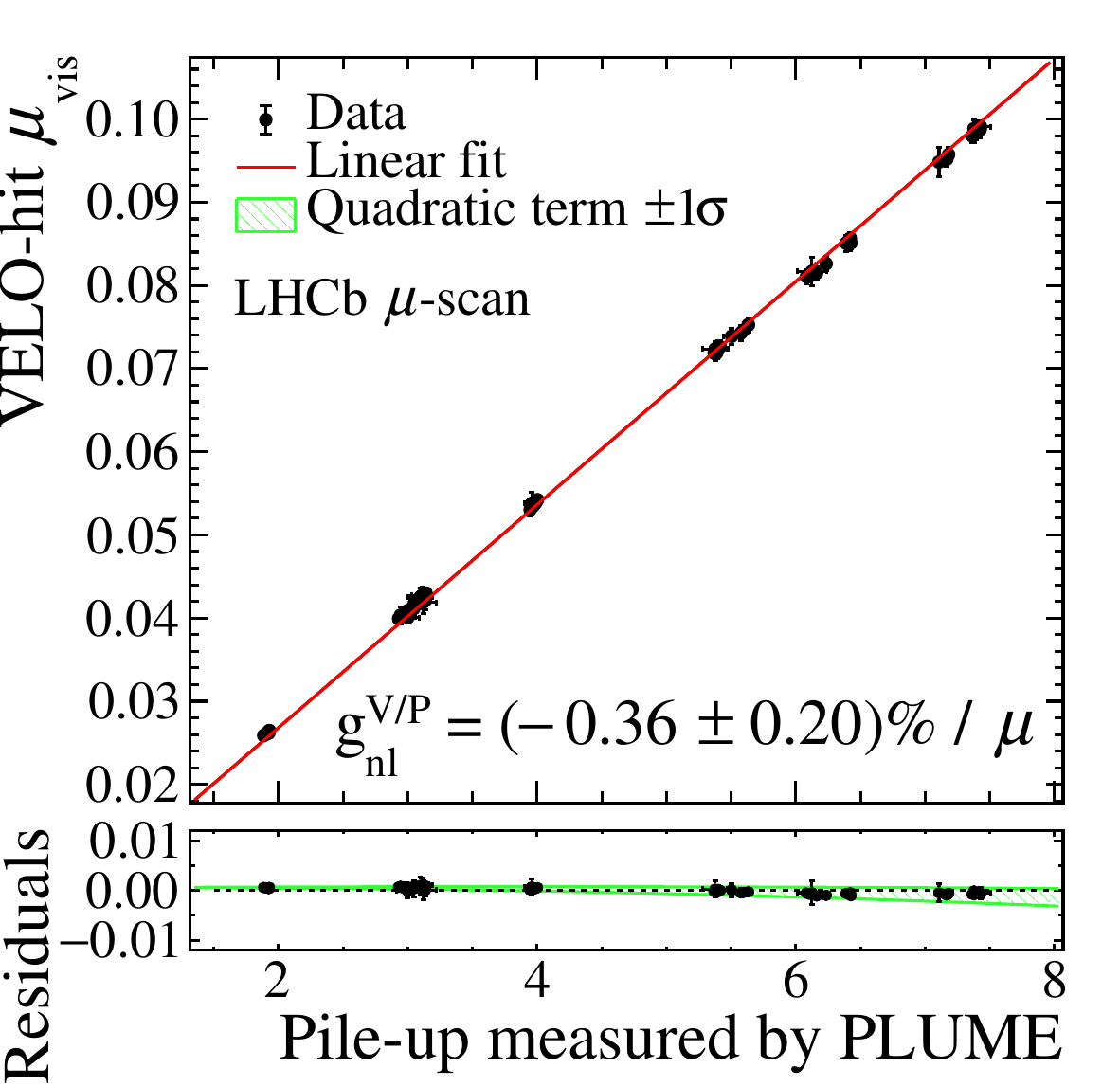}
    \caption{Analysis of the linearity of the \muvis estimators based on VELO cluster counting. The green bands in the residuals plots show the contribution of the quadratic coefficient evaluated at its central value, as obtained from the fits, plus/minus one standard deviation. (Left) Estimated \muvis versus true number of visible interaction, for simulated samples in Run~3 conditions. (Centre) Linear relation between the background-subtracted average-method and log0-method estimators from the same accumulation region during a $\mu$-scan performed in 2024. (Right) Linear relation between VELO-hit and PLUME measurements during the same $\mu$-scan.}
    \label{fig:mu-scans}
\end{figure}

Thanks to the high granularity of the VELO detector, the occupancy on its sensors is expected to have good linearity with respect to the instantaneous luminosity.
An intrinsic level of nonlinearity is expected in high-occupancy scenarios due to the superposition of hits, that cannot be reconstructed as separate clusters if they are not separated by at least one VELO pixel. A small saturation effect is therefore expected.
On the other hand, due to resource usage constraints, the clustering algorithm used on the VELO readout boards imposes limitations such as a maximum number of pixels per cluster, and a maximum number of clustering instances~\cite{Bassi:2023jpv}. These approximations can make large clusters partially reconstructed, or split. In the same way, clusters may be split when the allocated firmware resources are not sufficient to process very crowded events.
Cluster splitting can also lead to nonlinear effects, towards an excess of reconstructed hits. 

The linearity of the hit accumulators has been studied in simulation for an average number of concurrent \proton\proton interactions visible in LHCb (pile-up) up to $\mu= 25$, as well as in collision data using a dedicated calibration run, called $\mu$-scan, where the pile-up was varied up to $\mu = 7.7$.
Figure~\ref{fig:mu-scans} shows the linearity of the luminosity counters considered in this paper. The left-hand side panel uses simulation to assess the linearity of the background-subtracted \muvis, as measured by one of the average-method counters, with respect to the number of \proton\proton collisions per event. The level of nonlinearity is quantified in simulation by the parameter \mbox{$g_{\rm nl}^{\rm sim} (\mu) = (0.11\pm 0.05)\% / \mu$}, evaluated by fitting the measured \vs generated $\muvis$ relation with a second-order polynomial function, and evaluating the ratio of the quadratic term coefficient to the linear one.
It is not possible to perform the same study on collision data, as one has no access to the ``true'' visible number of collisions. One can, however, compare various estimators for \muvis against one another. As discussed in Sect.~\ref{sec:lumi_intro}, the log0 counters are expected to be insensitive to saturation effects, and are therefore used to this aim.
The central panel of Fig.~\ref{fig:mu-scans} compares the average and log0 counters from the same accumulation region, using collision data from a $\mu$-scan. This comparison confirms the linearity of the average counter, with a relation \mbox{$\muvis^{\rm avg} = \left(0.998\pm 0.002\right)\muvis^{\rm log0}$}. The nonlinearity coefficient is found to be compatible with zero: \mbox{$g_{\rm nl}^{\rm avg} = \left(0.02 \pm 0.22\right) / \mu_{\rm vis}$}. For the maximum value of $\mu_{\rm vis}$ recorded during the $\mu$-scan, corresponding to a pile-up of approximately 7.7 (significantly higher than the nominal value of 5.3 for LHCb Run~3), the relative difference between a quadratic and a linear regression amounts to $\left(0.3 \pm 2.0\right)\%$. These results are general and their validity extends to all of the accumulation regions.
Finally, Fig.~\ref{fig:mu-scans} (right) compares one VELO-hit estimator to the number of interactions recorded by PLUME, using data from the same $\mu$-scan. A linear relation is found, with a nonlinearity coefficient of $g_{\rm nl}^{\rm V/P}(\mu) = \left(-0.36\pm 0.20\right)\% / \mu$. 
This results in a maximum quadratic \vs linear regression difference of $(1.0\pm 1.3)\%$ at $\mu=7.7$.
These results show the good linear behaviour of the luminosity counters presented in this paper.

\subsection{Time stability}

\begin{figure}[b]
    \centering
    \raisebox{-.5\height}{\includegraphics[width=0.64\linewidth]{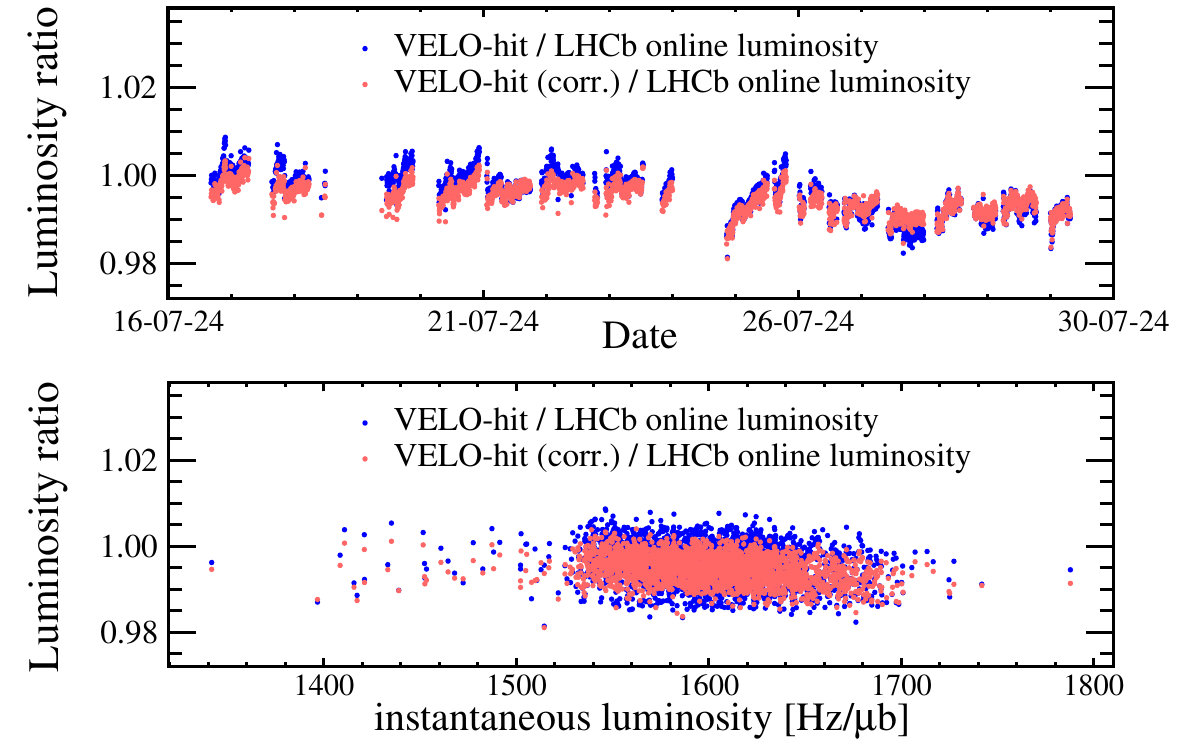}}\raisebox{-.5\height}{\includegraphics[width=0.35\linewidth]{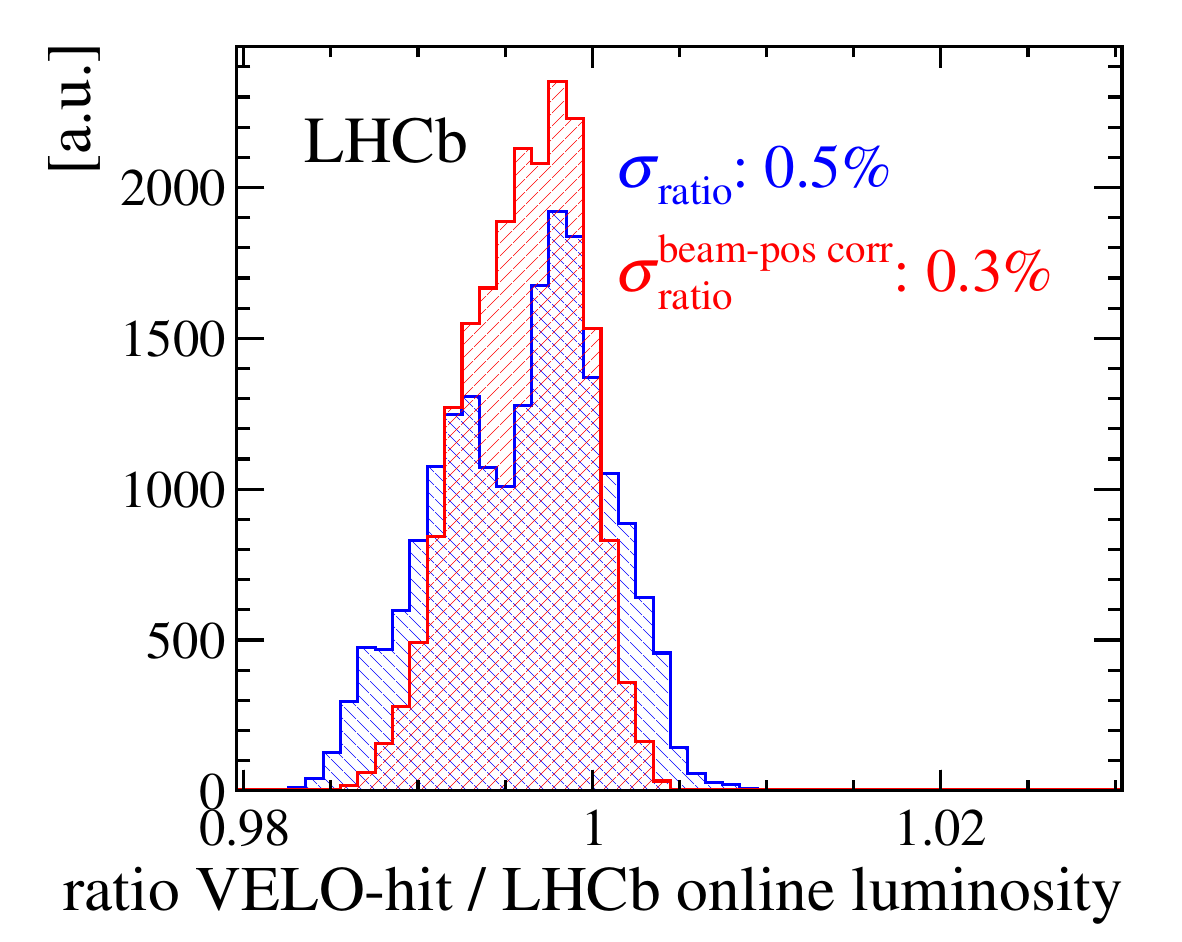}}
    \caption{Stability of the VELO-hit-based luminosity estimator with respect to the LHCb online luminosity over a two-weeks period in July 2024. The absolute value should not be taken as final, as it relies on fine calibration of the two luminosity measurements that has not been performed for this work. Data are shown before (blue) and after (red) applying the beam-spot position correction. (Left) Ratio between the luminosity estimators shown (top panel) as a function of time  and  (bottom panel) as a function of the measured instantaneous luminosity.
    (Right) Distribution of the ratio between the VELO-hit estimator and the LHCb online luminosity measurements, before and after correction. The effect of the beam-position correction is evident.
        }
    \label{fig:ratios-plume-bran}
\end{figure}

Stability throughout data taking is a crucial feature for a luminosity counter. The stability of the VELO-hit based counters is evaluated by comparing them to the LHCb online luminosity, which is mainly provided by the PLUME detector~\cite{LHCb-TDR-022}.

Data from two weeks of operation during July 2024 are used as a test sample to assess the stability of the counters. During these two weeks, luminosity varied approximately between 1500 and 1700~\!\hz/\!\mub. Figure~\ref{fig:ratios-plume-bran} shows the ratio of the luminosity measured by the global VELO-hit based estimator with respect to the LHCb online luminosity, throughout the whole period. The data exhibit remarkable stability in time. The spread of the ratio across the 2 weeks of operations is 0.5\%, and it shrinks to 0.3\% when systematic effects due to the position of the luminous region are corrected for (see Sect.~\ref{sec:systs}).
By fitting the distribution of the ratio as a function of luminosity, shown in the bottom panel of Fig.~\ref{fig:ratios-plume-bran} (left), with a first order polynomial function, a deviation from collinearity of about $\left(2.2\pm 0.2\right)\times 10^{-5}~{\left(\!\hz/\!\mub\right)}^{-1}$ is measured between the two luminometers, resulting in a maximum variation of $\left(0.43\pm0.03\right)$\% between the lower and upper boundaries of the tested luminosity range. This variation is less than half of the variation range of the ratio itself, and thus it is not explored further.

\subsection{Background subtraction and sensitivity to gas injection}
\begin{figure}
    \centering
    \includegraphics[width=0.46\linewidth]{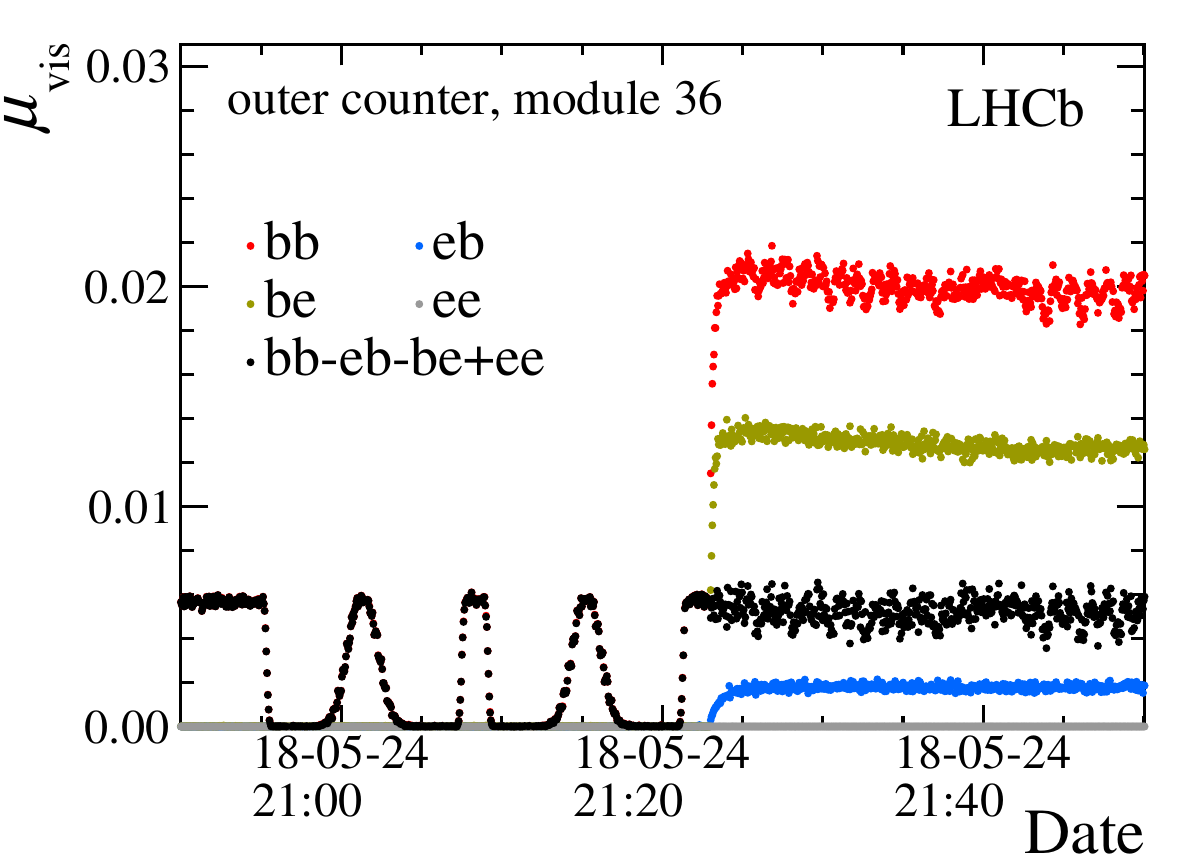}
    \includegraphics[width=0.47\linewidth]{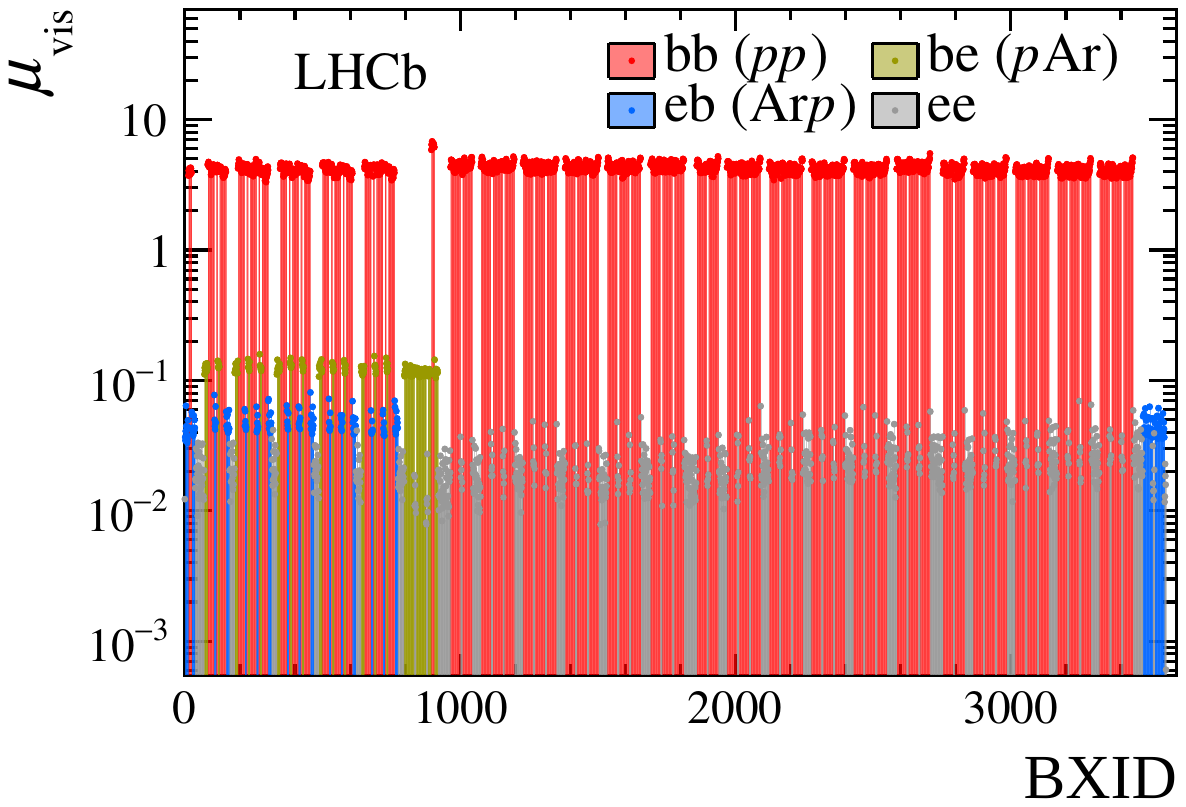}
    \caption{(Left) Number of visible interactions per bunch-crossing type, for one counter, during a part of the 2024 $pp$ vdM programme when argon was injected in the SMOG2 storage cell around 21:25. The data series in black represents the background-subtracted counter. The wider spread in the bb contribution,  as well as the \muvis slope in the bb and be ones, are due to the movements of the beam during the LSC scan. (Right) Per-BXID measurement of $\muvis$ during a short period of $pp$ collisions in 2024, averaged with a trimmed mean over all counters as done for the average global estimator.}
    \label{fig:smog_bkg_subtraction}
\end{figure}

During physics data taking, various gases are injected upstream of the VELO in the SMOG2 cell, allowing for a wide fixed-target physics programme~\cite{LHCb-DP-2024-002}.
Collisions of beam particles with gas can modify the hit count distribution. These contributions are removed by the background subtraction formulas from Sect.~\ref{sec:counters}, that make use of separate counters for different collision types.
After background subtraction, a slightly larger spread in the luminosity measurements is expected with respect to data collected without gas due to 
statistical fluctuations of the per-collision-type counters.
Figure~\ref{fig:smog_bkg_subtraction} (left) shows the effect of argon injection on the counters during the 2024 $pp$ vdM programme, demonstrating the effectiveness of the background-subtraction procedure. The right-hand side panel shows the measured luminosity as a function of the bunch-crossing identifier during a LHC fill in physics conditions, with argon injected in the SMOG2 cell. The $\rm be$, $\rm eb$, and $\rm ee$ bunch crossings have nonzero luminosities, which include contributions from $p$Ar and Ar$p$ interactions and beam- and detector-related backgrounds, respectively.

\subsection{Internal and external consistency}

The internal consistency of the hit-based luminosity estimator is assessed by considering the spread of the luminosities measured by the available counters. The left-hand side panel of Fig.~\ref{fig:trim} shows the values measured by all counters for a short period of time during physics data taking.
The average spread of all the average-method counters during the two-week long period considered for this study is $\sim$2.2\%, which reduces to about 1\% when the 15\% lowest and highest values are discarded, as is done to calculate the value of the trimmed-mean global estimator. The spread of all luminosity counters during the $pp$ vdM programme of 2024 was $0.8\%$ (Fig.~\ref{fig:vdm24pp}, right-hand side panel). The increased spread during physics data-taking is not fully understood, but it can be due to various systematic effects, such as beam-related biases, as discussed above, but also possibly intrinsic detector-related effects and residual nonlinearities. More details are discussed in Sect.~\ref{sec:systs}.

\begin{figure}[tb]
    \centering
    \includegraphics[width=0.8\linewidth]{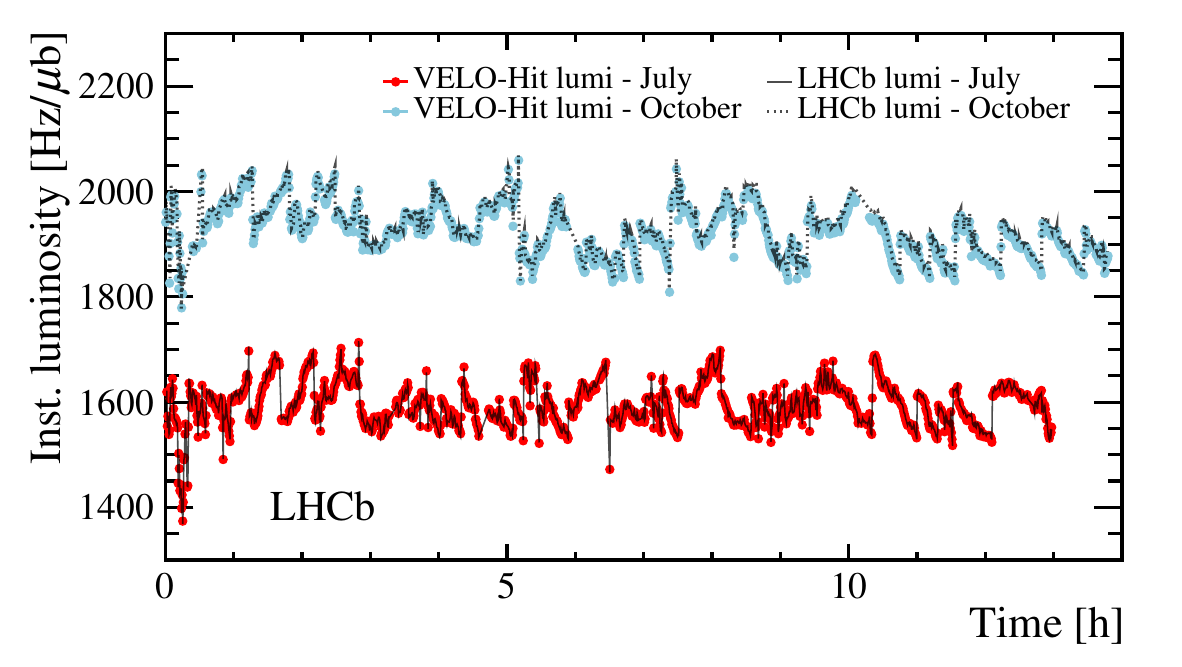}
    \caption{Instantaneous luminosities as measured by the global VELO-hit estimator and by PLUME during a 13-hour \proton\proton LHC fill in July 2024, randomly picked among the dataset analysed in this paper, as well as during a 14-hour fill at target Run~3 luminosity recorded in October 2024. The $x$ axis reports the time since the beginning of physics data taking in the fill. The sawtooth trend is due to the luminosity levelling procedure.
    While the absolute scale of the measured luminosity can only be validated with further analysis, the overlay highlights that the two measurements show the same trends.
    }
    \label{fig:retina_plume_comparison}
\end{figure}

Figure~\ref{fig:retina_plume_comparison} shows a comparison between the LHCb online luminosity and the global luminosity estimator based on the VELO hits, using data recorded during two \proton\proton fills at different luminosities
in July and October 2024. 
The luminosity levelling steps, as well as the other luminosity fluctuations, can be precisely tracked by both luminosity estimators, which remain consistent with each other throughout the fill.

\subsection{Lead-Lead luminosity measurement}

The LHCb collaboration performs measurements using both $pp$ and heavy-ion collisions. For the latter programme, lead ions are collided at a centre-of-mass energy of \mbox{$\sqsnn=5.36\tev$}. Just as in the \proton\proton case, a precise real-time monitoring of the instantaneous luminosity is required in order to operate the LHCb detector efficiently. 
The PbPb programme offers a unique opportunity to assess the validity and robustness of the hit-counting method to measure the instantaneous luminosity in very high-occupancy scenarios, dramatically more crowded than the average \proton\proton collision. In fact, PbPb collisions can have up to $\sim$70 times the amount of clusters found in a \proton\proton collision.
The FPGA clustering algorithm used in \proton\proton physics suffers from the high occupancy of PbPb collisions. In order to provide good reconstruction quality in heavy-ion conditions, the algorithm is optimised to reduce the number of split clusters, improving linearity in very busy events.
This is achieved by exploiting the low interaction rate of PbPb collisions, $\mathcal{O}\left(10\khz\right)$, in order to dynamically allocate more FPGA resources for the handling of busy events.

\begin{figure}
    \centering
    \raisebox{-.5\height}{\includegraphics[width=.45\linewidth]{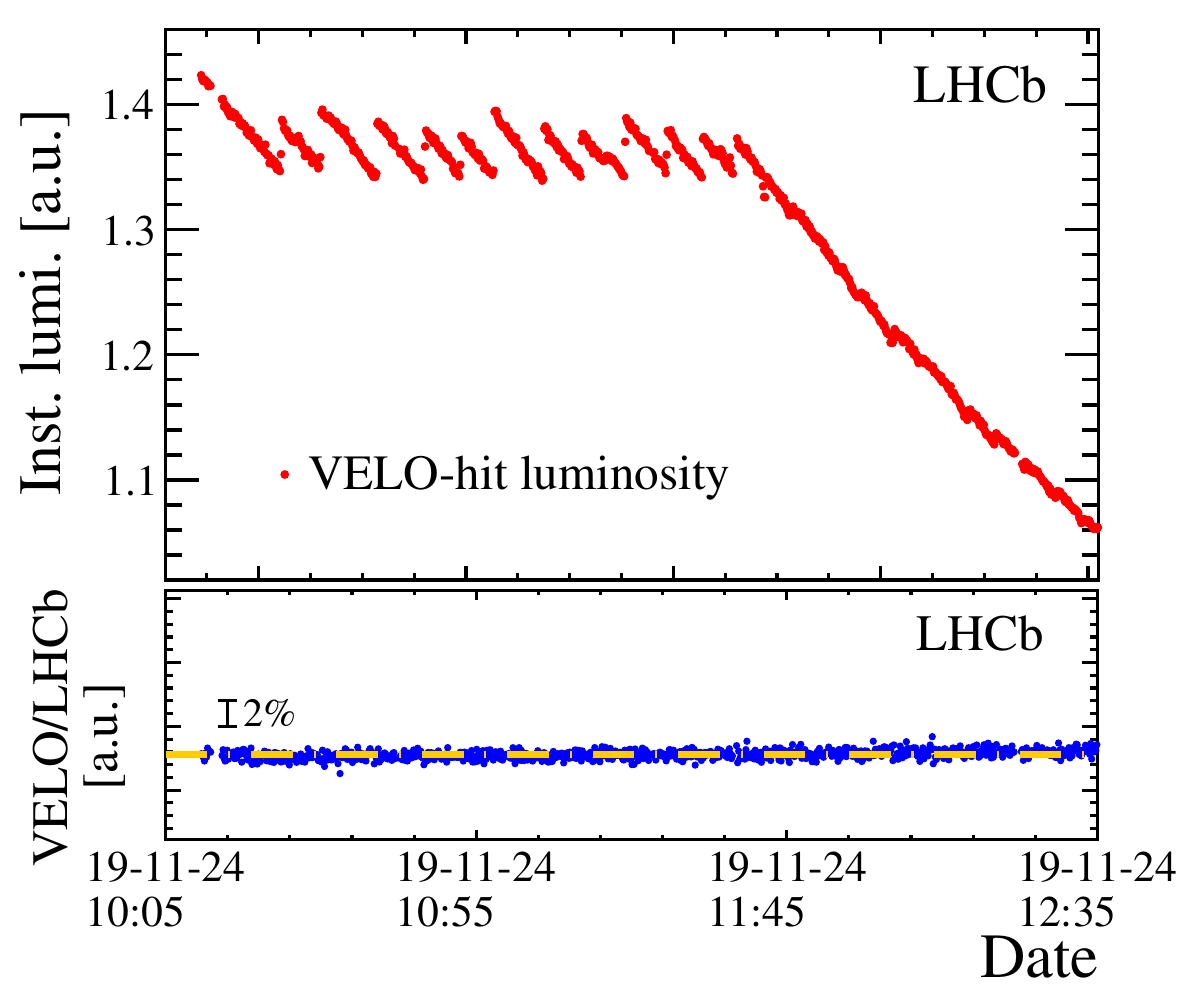}}
    \raisebox{-.5\height}{\includegraphics[width=.52\linewidth]{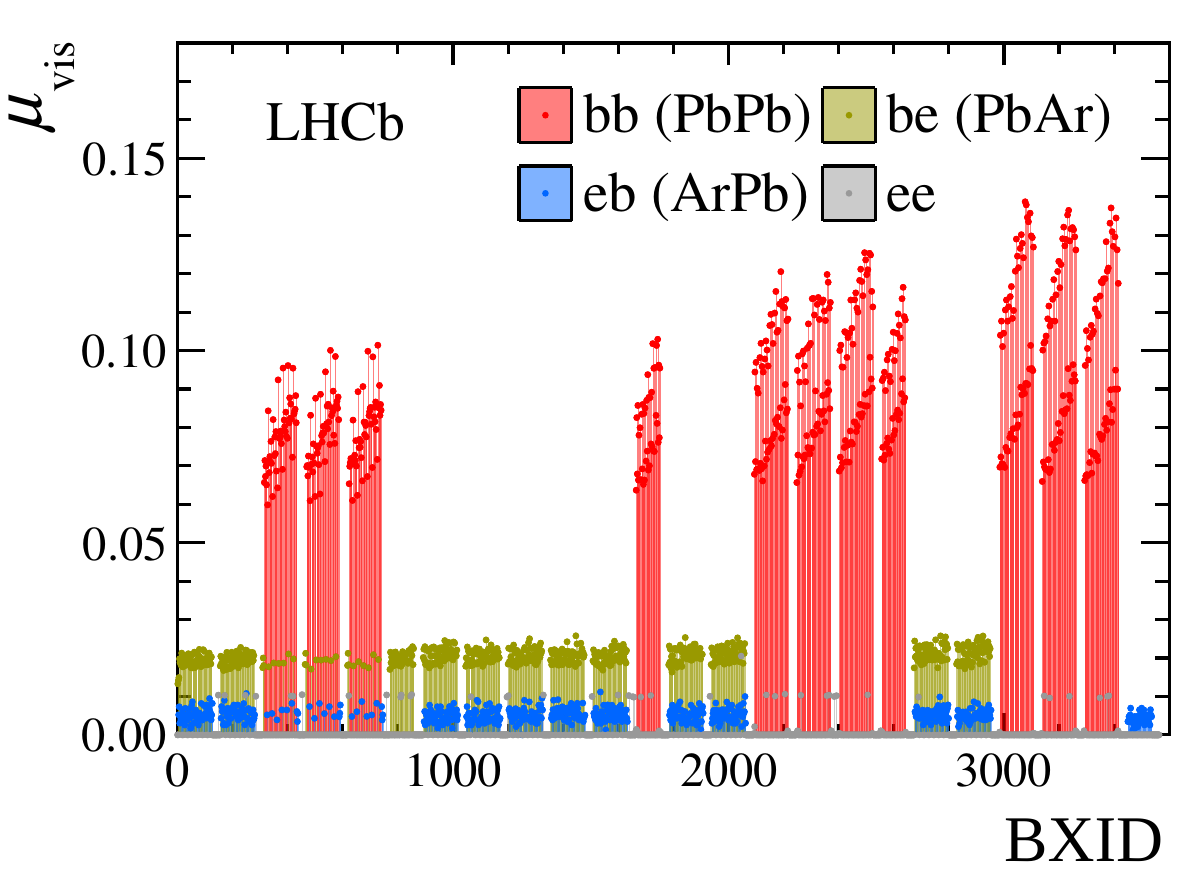}}
    \caption{Performance during the PbPb programme. (Left) Instantaneous luminosity measured by the global VELO-hit estimator during a PbPb fill in 2024 without any gas injected in the SMOG2 cell. The ratio with respect to the LHCb online luminosity is also shown, with the purpose of assessing stability in time. (Right) Per-BXID \muvis measurements during a fill with argon injected in the SMOG2 cell. A significant variation of the beam-beam luminosity estimates is observed.}
    \label{fig:lumi_PbPb}
\end{figure}

Hit counters are expected to saturate for high-occupancy events due to the physical overlap of particle tracks on the VELO sensors. In principle, the log0 method is expected to be better suited for its robustness against saturation effects.
However, due to the very low interaction rate of PbPb collisions ($\mu_{\text{PbPb}}\!\sim\!10^{-3}$), it can be demonstrated that also the average-method counters are still linear with respect to the instantaneous luminosity.
The probability of more than one PbPb collision occurring in the same bunch-crossing is extremely low, of the order of $10^{-6}$. Therefore,  saturation in nonempty events only depends on the intrinsic hit distribution of PbPb collisions, which is itself independent of the luminosity. Any saturation effect in the hit distribution is accounted for in the calibration phase, resulting in an average method-based estimator that is with good approximation linear with respect to the instantaneous luminosity.\footnote{Specifically, \muvis can be expressed as an expansion where terms beyond the linear one are suppressed by a factor $\mu$, which in PbPb collisions is $\mathcal{O}(10^{-3})$.} As in the $pp$ case, the average method is preferred over log0 due to the nonuniformity of the bunch populations, which is clearly visible in the right-hand side panel of Fig.~\ref{fig:lumi_PbPb}. The left-hand side panel of Fig.~\ref{fig:lumi_PbPb} shows the trend of the VELO-hit based instantaneous luminosity estimator during a PbPb fill without gas injection. The ratio between the VELO-hit based measurement and the LHCb online one is stable in time throughout the fill, within 2$\%$. A quantitative study of the PbPb luminosity requires further work both on these counters and on a reference measurement, and it is beyond the scope of the present work, but the potential of this new method is clear.

\subsection{Sensitivity to systematic effects}
\label{sec:systs}
The distribution of hits on the VELO detector, and thus the related luminosity measurement, can be affected by several factors, that can be grouped in two main categories. 
\begin{itemize}
    \item Sensor-related effects: local inefficiencies, ``hot'' pixels and noisy regions of the silicon substrate. Ageing of the VELO sensors is not taken into account for the results presented in this paper, which are based on only one year of operation, during which the local inefficiencies have been very low. However, in the long term, it may have a non-negligible impact in the determination of the luminosity~\cite{CMS:2021xjt}. Effects of radiation damage on the silicon bulk are monitored during special scans in which the hit efficiency is evaluated as a function of the bias voltage~\cite{Rinnert:2015uns}.
    \item Beam-related effects: shifts and changes in the shape of the luminous region intrinsically vary the hit distribution on the VELO sensors.
\end{itemize}
The net result of these experimental effects must be taken into account, and used to either correct the luminosity measurement, or infer a systematic uncertainty. Thanks to the use of the trimmed mean, sensor-related effects are expected to be suppressed: outlier luminosity counters are dynamically excluded from the computation of the global luminosity estimator.

\begin{figure}
    \centering
    \includegraphics[width=.328\linewidth]{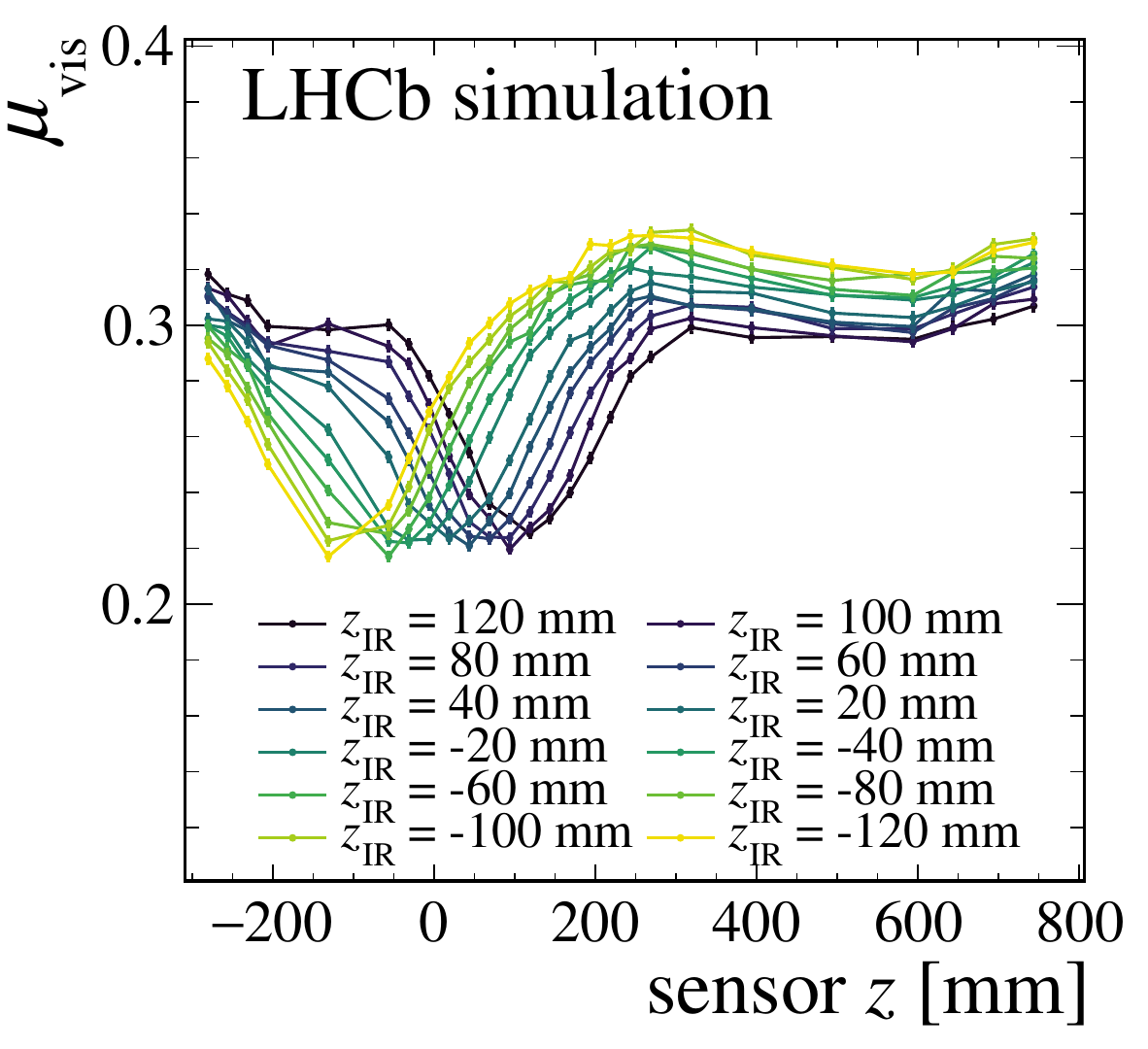}\hfill
    \includegraphics[width=.328\linewidth]{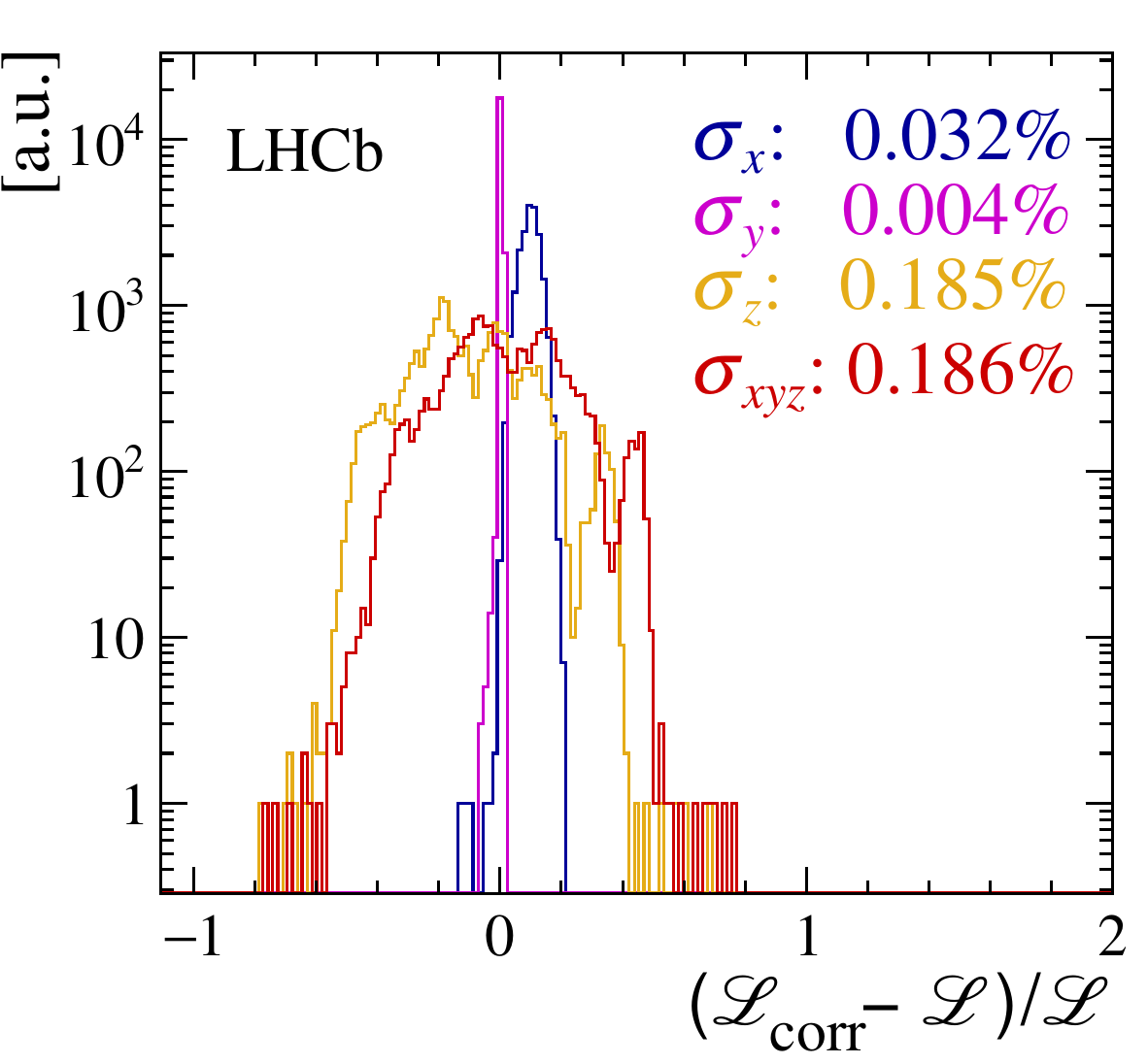}\hfill
    \includegraphics[width=.328\linewidth]{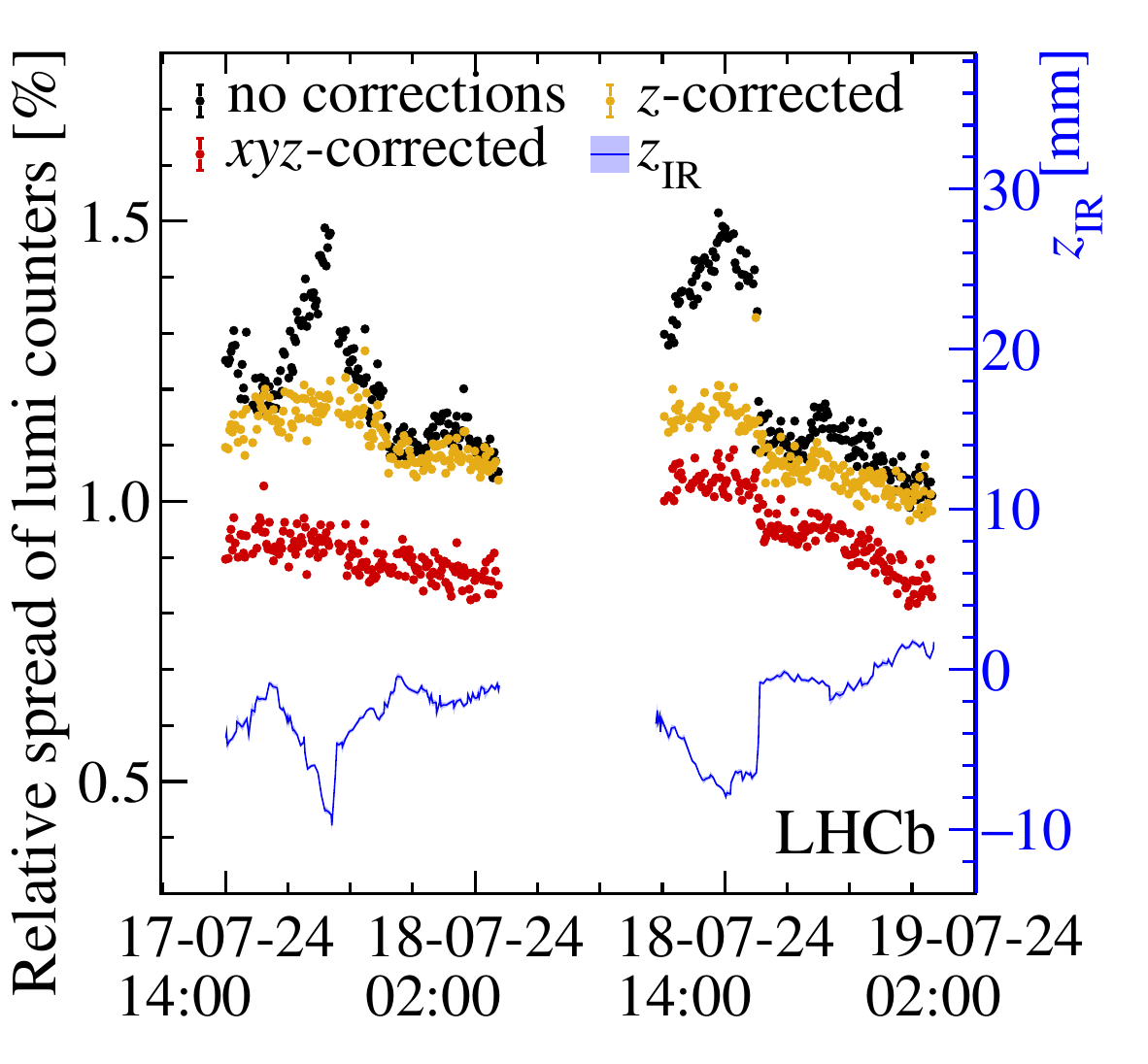}
    \caption{(Left) Simulated distribution of \muvis as a function of the sensor $z$ position. Each colour represents a simulated sample with different position of the interaction region along the $z$ direction~\cite{Passaro:2842603}. Across the whole 2024 $pp$ data-taking period, the centroid of the interaction region has been observed to fluctuate within a $\pm 20\mm$ $z$ range. (Centre) Distribution of the relative difference between the position-corrected luminosity estimates and the uncorrected ones, for a two-week dataset collected in July 2024. The $\sigma_{i}$ values listed in the plot are the standard deviations of the relative luminosity variations observed when the $i$-axis position correction is applied. (Right) Trend of the spread of the luminosity counters contributing to the trimmed-mean estimator for a \SI{36}{\hour} period in July 2024. For reference, the $z$ position of the interaction region centroid is also shown, as measured by the LHCb track-based monitoring tasks. 
    }
    \label{fig:sys_xyz}
\end{figure}

In the context of this work, only beam-related effects are explicitly discussed.
Studies based on simulated data~\cite{Passaro:2842603} have shown that the distribution of hits on the VELO sensors depends on the position and size of the luminous region, and it is most sensitive to shifts along the $z$ direction, as visible in the left-hand side panel of Fig.~\ref{fig:sys_xyz}. In order to measure the effect of such movements on the luminosity estimate, the test sample from Fig.~\ref{fig:ratios-plume-bran} is reanalysed. Each luminosity counter is corrected for the shift of the luminous region with respect to the position it had during the calibration vdM run. The correction is modelled using simulated samples with different beam-spot positions. A correction factor $c_i^k$ is computed for each luminosity counter and each Cartesian component $x_i$ of the beam-spot position as $c_i^k = 1-f_i^k(x_i)$, where $x_i=\{x,y,z\}$ and $f_i^k$ is the relative variation of the counter $k$ with respect to its value  when the interaction region is in the nominal position. The final correction for counter $k$ is given by $c_x^kc_y^kc_z^k$. The trimmed-mean estimator is recomputed offline and it is compared to its uncorrected online value to infer the magnitude of the effect. The results of this analysis are gathered in Fig.~\ref{fig:sys_xyz}, which shows that, on average, the luminosity measurement is affected by a $0.2\%$ systematic effect. By far the most relevant contribution (0.185\%, see central panel of Fig.~\ref{fig:sys_xyz}) stems from the displacement of the interaction region along the $z$ direction. The beam-position correction is effective in correcting the luminosity counters estimates: the right panel of Fig.~\ref{fig:sys_xyz} shows that the relative spread of the counters shrinks once the correction is applied, proving the consistency of the method. 
In fact, correcting the counters for the actual $z$ position of the luminous region decouples the spread of the counters from the position movements. Further correcting for the $x$ and $y$ positions reduces the spread down to about 1\%, from peaks of up to 1.5\%.
While this plot shows a period during which the $z$ position of the luminous region varies visibly, the spread of the position-corrected counters is found to remain stable at the $\sim$1\% level throughout the two-week dataset shown in the central panel.

\subsection{Availability and downtime}
During a typical \proton\proton LHC fill, the availability of the hit-count based luminosity measurement has been as high as $\sim$93\% of the stable-beams time. The system is normally online whenever the LHCb DAQ is active. When it is not, the ECS forces a temporary DAQ stop, which in turn triggers a halt of the hit processing sequence in the VELO readout boards.
A small source of inefficiency ($\sim$1\%) stems from the time needed to close the VELO to its nominal beam-centred position at the start of every fill. In this case, hit processing is enabled, but the acceptance of each accumulation region changes while the VELO is moving. As a consequence, the calibration performed during the vdM scan ---during which the VELO is closed--- is not valid in this phase. Another $\sim$2\% inefficiency is due to self-consistency checks based on real-time analysis of the distribution of the luminosity measurements from all the counters. These checks have the purpose to determine whether the system is in stable conditions and evaluate the possible presence of outliers.

\section{Conclusions}
\label{sec:concl}

Luminosity determination is a crucial measurement for any high-energy physics experiment. In LHCb, a precise real-time instantaneous luminosity measurement is needed in order to operate the experiment in stable, efficient and safe conditions.

In this paper, a new method to monitor the instantaneous luminosity in real time, based on counters of reconstructed VELO hits, is presented. The development of this method was enabled by the implementation of a fast FPGA algorithm capable of reconstructing two-dimensional hits on the VELO sensors already at  the readout level.
Hit statistics, accumulated directly on the readout FPGAs, are calibrated with van der Meer scans in order to obtain luminosity measurements. These measurements are combined in a trimmed-mean-based luminosity estimator, robust against random fluctuations of the hit counters, and against failing or noisy channels. The estimator is available in real time, with a granularity dominated by the desired frequency of readings of the FPGA registers. Throughout the 2024 data-taking period, updated counter values were made available every 90\ms, and read out every 3\sec via ECS.

As discussed in Sect.~\ref{sec:results}, the luminosity estimator exhibits:
\begin{itemize}
    \item excellent linear behaviour, with saturation effects compatible with zero;
    \item very good stability, with a relative spread against the LHCb online luminosity of 0.3\% over two weeks of operation (once beam-position corrections are applied);
    \item high availability, with as low as 2\% intrinsic inefficiency.
\end{itemize}

Within the counters dynamically chosen to construct the global estimator, internal consistency is at the level of about 1\% in a two-week long period during physics data taking in $pp$ collisions. With the statistical uncertainty being negligible, the main contributions to the luminosity resolution are systematic effects due to:
\begin{itemize}
\item occupancy variation over the VELO sensors induced by geometrical movements of the interaction region: the correction has $\mathcal{O}(0.2\%)$ effect on the luminosity measurement;
\item systematic uncertainties related to the calibration procedure, which are not evaluated in this paper (the treatment of such systematics is thoroughly discussed in the literature~\cite{LHCb-PAPER-2014-047});
\item residual nonlinearity, which is compatible with zero when evaluated on simulation and on data taken during dedicated $\mu$-scans; a discrepancy of $\sim$0.4\% is however observed in a 2-week \proton\proton collision dataset between this luminometer and the LHCb official one.

\end{itemize}

The VELO hit-based instantaneous luminosity measurement has been operational during LHCb data taking throughout the 2024 run, and it remains in use to date. It shows excellent agreement with the LHCb online measurement, and the uncertainties are well within the limits required by the luminosity levelling procedure.
The results obtained in PbPb collisions prove that this method is valuable also in very high-occupancy scenarios.

Another unique characteristics of this method is its flexibility, stemming from the inherent customizability of the firmware implementation of the accumulation regions. These can be further tuned to achieve sensitivity in a range of data-taking conditions, and also to measure other quantities. Indeed, as a further application, the VELO hit counters can also be used to evaluate the displacements of the interaction region centroid in real time. Such a tool then allows one to calculate a position-corrected luminosity value in real time, as opposed to the method discussed above. This measurement can be performed by constructing appropriate statistics sensitive to the beam-spot position.
A method based on principal component analysis, relating counter rates to the spatial coordinates of the luminous region once calibrated on simulation, is currently being commissioned, having shown promising results~\cite{cordova2025}.

\section*{Acknowledgements}
%
%
\noindent We express our gratitude to our colleagues in the CERN
accelerator departments for the excellent performance of the LHC. We
thank the technical and administrative staff at the LHCb
institutes.
We acknowledge support from CERN and from the national agencies:
ARC (Australia);
CAPES, CNPq, FAPERJ and FINEP (Brazil); 
MOST and NSFC (China); 
CNRS/IN2P3 (France); 
BMFTR, DFG and MPG (Germany);
INFN (Italy); 
NWO (Netherlands); 
MNiSW and NCN (Poland); 
MCID/IFA (Romania); 
MICIU and AEI (Spain);
SNSF and SER (Switzerland); 
NASU (Ukraine); 
STFC (United Kingdom); 
DOE NP and NSF (USA).
We acknowledge the computing resources that are provided by ARDC (Australia), 
CBPF (Brazil),
CERN, 
IHEP and LZU (China),
IN2P3 (France), 
KIT and DESY (Germany), 
INFN (Italy), 
SURF (Netherlands),
Polish WLCG (Poland),
IFIN-HH (Romania), 
PIC (Spain), CSCS (Switzerland), 
and GridPP (United Kingdom).
We are indebted to the communities behind the multiple open-source
software packages on which we depend.
Individual groups or members have received support from
Key Research Program of Frontier Sciences of CAS, CAS PIFI, CAS CCEPP, 
Minciencias (Colombia);
EPLANET, Marie Sk\l{}odowska-Curie Actions, ERC and NextGenerationEU (European Union);
A*MIDEX, ANR, IPhU and Labex P2IO, and R\'{e}gion Auvergne-Rh\^{o}ne-Alpes (France);
Alexander-von-Humboldt Foundation (Germany);
ICSC (Italy); 
Severo Ochoa and Mar\'ia de Maeztu Units of Excellence, GVA, XuntaGal, GENCAT, InTalent-Inditex and Prog.~Atracci\'on Talento CM (Spain);
SRC (Sweden);
the Leverhulme Trust, the Royal Society and UKRI (United Kingdom).

\addcontentsline{toc}{section}{References}
\bibliographystyle{LHCb}
\bibliography{main,standard,LHCb-PAPER,LHCb-CONF,LHCb-DP,LHCb-TDR}

\ifx\mcitethebibliography\mciteundefinedmacro
\PackageError{LHCb.bst}{mciteplus.sty has not been loaded}
{This bibstyle requires the use of the mciteplus package.}\fi
\providecommand{\href}[2]{#2}
\begin{mcitethebibliography}{10}
\mciteSetBstSublistMode{n}
\mciteSetBstMaxWidthForm{subitem}{\alph{mcitesubitemcount})}
\mciteSetBstSublistLabelBeginEnd{\mcitemaxwidthsubitemform\space}
{\relax}{\relax}

\bibitem{Bruning:2004ej}
O.~S. Brüning {\em et~al.}, \ifthenelse{\boolean{articletitles}}{\emph{{LHC design report Vol.1: The LHC main ring}}}{}  CERN-2004-003-V1, CERN-2004-003, CERN-2004-003-V-1, CERN, 2004.
\newblock doi:~\href{https://doi.org/10.5170/CERN-2004-003-V-1}{10.5170/CERN-2004-003-V-1}\relax
\mciteBstWouldAddEndPuncttrue
\mciteSetBstMidEndSepPunct{\mcitedefaultmidpunct}
{\mcitedefaultendpunct}{\mcitedefaultseppunct}\relax
\EndOfBibitem
\bibitem{LHCb-DP-2008-001}
LHCb collaboration, A.~A. Alves~Jr.\ {\em et~al.}, \ifthenelse{\boolean{articletitles}}{\emph{{The \lhcb detector at the LHC}}, }{}\href{https://doi.org/10.1088/1748-0221/3/08/S08005}{JINST \textbf{3} (2008) S08005}\relax
\mciteBstWouldAddEndPuncttrue
\mciteSetBstMidEndSepPunct{\mcitedefaultmidpunct}
{\mcitedefaultendpunct}{\mcitedefaultseppunct}\relax
\EndOfBibitem
\bibitem{Follin:2014nva}
F.~Follin and D.~Jacquet, \ifthenelse{\boolean{articletitles}}{\emph{{Implementation and experience with luminosity levelling with offset beam}}, }{} in {\em {ICFA Mini-Workshop on Beam-Beam Effects in Hadron Colliders}}, \href{https://doi.org/10.5170/CERN-2014-004.183}{ 183--187, 2014}, \href{http://arxiv.org/abs/1410.3667}{{\normalfont\ttfamily arXiv:1410.3667}}\relax
\mciteBstWouldAddEndPuncttrue
\mciteSetBstMidEndSepPunct{\mcitedefaultmidpunct}
{\mcitedefaultendpunct}{\mcitedefaultseppunct}\relax
\EndOfBibitem
\bibitem{LHCb-TDR-022}
LHCb collaboration, \ifthenelse{\boolean{articletitles}}{\emph{{LHCb PLUME: Probe for LUminosity MEasurement}}, }{} \href{http://cdsweb.cern.ch/search?p=CERN-LHCC-2021-002&f=reportnumber&action_search=Search&c=LHCb} {CERN-LHCC-2021-002}, 2021\relax
\mciteBstWouldAddEndPuncttrue
\mciteSetBstMidEndSepPunct{\mcitedefaultmidpunct}
{\mcitedefaultendpunct}{\mcitedefaultseppunct}\relax
\EndOfBibitem
\bibitem{Bassi:2023jpv}
G.~Bassi {\em et~al.}, \ifthenelse{\boolean{articletitles}}{\emph{{A FPGA-based architecture for real-time cluster finding in the LHCb silicon pixel detector}}, }{}\href{https://doi.org/10.1109/TNS.2023.3273600}{IEEE Trans.\ Nucl.\ Sci.\  \textbf{70} (2023) 1189}, \href{http://arxiv.org/abs/2302.03972}{{\normalfont\ttfamily arXiv:2302.03972}}\relax
\mciteBstWouldAddEndPuncttrue
\mciteSetBstMidEndSepPunct{\mcitedefaultmidpunct}
{\mcitedefaultendpunct}{\mcitedefaultseppunct}\relax
\EndOfBibitem
\bibitem{LHCb-DP-2022-002}
LHCb collaboration, R.~Aaij {\em et~al.}, \ifthenelse{\boolean{articletitles}}{\emph{{The LHCb Upgrade I}}, }{}\href{https://doi.org/10.1088/1748-0221/19/05/P05065}{{JINST} \textbf{19} (2024) P05065}, \href{http://arxiv.org/abs/2305.10515}{{\normalfont\ttfamily arXiv:2305.10515}}\relax
\mciteBstWouldAddEndPuncttrue
\mciteSetBstMidEndSepPunct{\mcitedefaultmidpunct}
{\mcitedefaultendpunct}{\mcitedefaultseppunct}\relax
\EndOfBibitem
\bibitem{LHCb-TDR-013}
LHCb collaboration, \ifthenelse{\boolean{articletitles}}{\emph{{LHCb VELO Upgrade Technical Design Report}}, }{} \href{http://cdsweb.cern.ch/search?p=CERN-LHCC-2013-021&f=reportnumber&action_search=Search&c=LHCb} {CERN-LHCC-2013-021}, 2013\relax
\mciteBstWouldAddEndPuncttrue
\mciteSetBstMidEndSepPunct{\mcitedefaultmidpunct}
{\mcitedefaultendpunct}{\mcitedefaultseppunct}\relax
\EndOfBibitem
\bibitem{LHCb-DP-2024-002}
O.~B. Garcia {\em et~al.}, \ifthenelse{\boolean{articletitles}}{\emph{{High-density gas target at the LHCb experiment}}, }{}\href{https://doi.org/https://doi.org/10.1103/PhysRevAccelBeams.27.111001}{Phys.\ Rev.\ Accel.\ Beams \textbf{27} (2024) 111001}, \href{http://arxiv.org/abs/2407.14200}{{\normalfont\ttfamily arXiv:2407.14200}}\relax
\mciteBstWouldAddEndPuncttrue
\mciteSetBstMidEndSepPunct{\mcitedefaultmidpunct}
{\mcitedefaultendpunct}{\mcitedefaultseppunct}\relax
\EndOfBibitem
\bibitem{Cachemiche_2016}
J.~P. Cachemiche {\em et~al.}, \ifthenelse{\boolean{articletitles}}{\emph{{The PCIe-based readout system for the LHCb experiment}}, }{}\href{https://doi.org/10.1088/1748-0221/11/02/P02013}{JINST \textbf{11} (2016) P02013}\relax
\mciteBstWouldAddEndPuncttrue
\mciteSetBstMidEndSepPunct{\mcitedefaultmidpunct}
{\mcitedefaultendpunct}{\mcitedefaultseppunct}\relax
\EndOfBibitem
\bibitem{arria10ds}
Intel Corporation, {\em Intel\textsuperscript{\textregistered} Arria\textsuperscript{\textregistered} 10 Device Datasheet}, 2022.
\newblock Available at \href{https://www.intel.com/content/www/us/en/docs/programmable/683771/current/device-datasheet.html}{\url{www.intel.com/content/www/us/en/docs/programmable/683771}}\relax
\mciteBstWouldAddEndPuncttrue
\mciteSetBstMidEndSepPunct{\mcitedefaultmidpunct}
{\mcitedefaultendpunct}{\mcitedefaultseppunct}\relax
\EndOfBibitem
\bibitem{Barbosa:2019vkc}
J.~Barbosa {\em et~al.}, \ifthenelse{\boolean{articletitles}}{\emph{{Front-end electronics control and monitoring for the LHCb Upgrade}}, }{}\href{https://doi.org/10.1051/epjconf/201921401002}{EPJ Web Conf.\  \textbf{214} (2019) 01002}\relax
\mciteBstWouldAddEndPuncttrue
\mciteSetBstMidEndSepPunct{\mcitedefaultmidpunct}
{\mcitedefaultendpunct}{\mcitedefaultseppunct}\relax
\EndOfBibitem
\bibitem{LHCB-TDR-018}
LHCb collaboration, \ifthenelse{\boolean{articletitles}}{\emph{{Computing Model of the Upgrade LHCb experiment}}, }{} \href{http://cdsweb.cern.ch/search?p=CERN-LHCC-2018-014&f=reportnumber&action_search=Search&c=LHCb} {CERN-LHCC-2018-014}, 2018\relax
\mciteBstWouldAddEndPuncttrue
\mciteSetBstMidEndSepPunct{\mcitedefaultmidpunct}
{\mcitedefaultendpunct}{\mcitedefaultseppunct}\relax
\EndOfBibitem
\bibitem{LHCB-TDR-017}
LHCb collaboration, \ifthenelse{\boolean{articletitles}}{\emph{{LHCb Upgrade Software and Computing}}, }{} \href{http://cdsweb.cern.ch/search?p=CERN-LHCC-2018-007&f=reportnumber&action_search=Search&c=LHCb} {CERN-LHCC-2018-007}, 2018\relax
\mciteBstWouldAddEndPuncttrue
\mciteSetBstMidEndSepPunct{\mcitedefaultmidpunct}
{\mcitedefaultendpunct}{\mcitedefaultseppunct}\relax
\EndOfBibitem
\bibitem{sym14050860}
P.~Albicocco {\em et~al.}, \ifthenelse{\boolean{articletitles}}{\emph{{A method based on muon system to monitor LHCb luminosity}}, }{}\href{https://doi.org/10.3390/sym14050860}{Symmetry \textbf{14} (2022) }\relax
\mciteBstWouldAddEndPuncttrue
\mciteSetBstMidEndSepPunct{\mcitedefaultmidpunct}
{\mcitedefaultendpunct}{\mcitedefaultseppunct}\relax
\EndOfBibitem
\bibitem{Pugatch_2025}
V.~Pugatch {\em et~al.}, \ifthenelse{\boolean{articletitles}}{\emph{{Metal foil detectors assembly for the beam and background monitoring in the LHCb experiment}}, }{}\href{https://doi.org/10.1088/1748-0221/20/07/P07027}{JINST \textbf{20} (2025) P07027}\relax
\mciteBstWouldAddEndPuncttrue
\mciteSetBstMidEndSepPunct{\mcitedefaultmidpunct}
{\mcitedefaultendpunct}{\mcitedefaultseppunct}\relax
\EndOfBibitem
\bibitem{LHCb-DP-2024-003}
P.~A. Cartelle {\em et~al.}, \ifthenelse{\boolean{articletitles}}{\emph{{Luminosity measurement with the LHCb RICH detectors in Run 3}}, }{}\href{http://arxiv.org/abs/2503.05273}{{\normalfont\ttfamily arXiv:2503.05273}}\relax
\mciteBstWouldAddEndPuncttrue
\mciteSetBstMidEndSepPunct{\mcitedefaultmidpunct}
{\mcitedefaultendpunct}{\mcitedefaultseppunct}\relax
\EndOfBibitem
\bibitem{Passaro:2024fkw}
D.~Passaro {\em et~al.}, \ifthenelse{\boolean{articletitles}}{\emph{{LHC beam monitoring via real-time hit reconstruction in the LHCb VELO pixel detector}}, }{}\href{http://arxiv.org/abs/2409.06524}{{\normalfont\ttfamily arXiv:2409.06524}}\relax
\mciteBstWouldAddEndPuncttrue
\mciteSetBstMidEndSepPunct{\mcitedefaultmidpunct}
{\mcitedefaultendpunct}{\mcitedefaultseppunct}\relax
\EndOfBibitem
\bibitem{Grafstrom:2015foa}
P.~Grafstr{\"o}m and W.~Kozanecki, \ifthenelse{\boolean{articletitles}}{\emph{{Luminosity determination at proton colliders}}, }{}\href{https://doi.org/10.1016/j.ppnp.2014.11.002}{Prog.\ Part.\ Nucl.\ Phys.\  \textbf{81} (2015) 97}\relax
\mciteBstWouldAddEndPuncttrue
\mciteSetBstMidEndSepPunct{\mcitedefaultmidpunct}
{\mcitedefaultendpunct}{\mcitedefaultseppunct}\relax
\EndOfBibitem
\bibitem{Zaitsev:2000js}
N.~Zaitsev, {\em {Study of the LHCb pile-up trigger and $B_s \rightarrow J/\psi \phi$ decay}}, PhD thesis, Amsterdam U., 2000, \url{https://cds.cern.ch/record/473383}\relax
\mciteBstWouldAddEndPuncttrue
\mciteSetBstMidEndSepPunct{\mcitedefaultmidpunct}
{\mcitedefaultendpunct}{\mcitedefaultseppunct}\relax
\EndOfBibitem
\bibitem{LHCb-PAPER-2014-047}
LHCb collaboration, R.~Aaij {\em et~al.}, \ifthenelse{\boolean{articletitles}}{\emph{{Precision luminosity measurements at LHCb}}, }{}\href{https://doi.org/10.1088/1748-0221/9/12/P12005}{JINST \textbf{9} (2014) P12005}, \href{http://arxiv.org/abs/1410.0149}{{\normalfont\ttfamily arXiv:1410.0149}}\relax
\mciteBstWouldAddEndPuncttrue
\mciteSetBstMidEndSepPunct{\mcitedefaultmidpunct}
{\mcitedefaultendpunct}{\mcitedefaultseppunct}\relax
\EndOfBibitem
\bibitem{CMS:2021xjt}
CMS collaboration, A.~M. Sirunyan {\em et~al.}, \ifthenelse{\boolean{articletitles}}{\emph{{Precision luminosity measurement in proton-proton collisions at $\sqrt{s} =$ 13 TeV in 2015 and 2016 at CMS}}, }{}\href{https://doi.org/10.1140/epjc/s10052-021-09538-2}{Eur.\ Phys.\ J.\  \textbf{C81} (2021) 800}, \href{http://arxiv.org/abs/2104.01927}{{\normalfont\ttfamily arXiv:2104.01927}}\relax
\mciteBstWouldAddEndPuncttrue
\mciteSetBstMidEndSepPunct{\mcitedefaultmidpunct}
{\mcitedefaultendpunct}{\mcitedefaultseppunct}\relax
\EndOfBibitem
\bibitem{Hennessy:2021aec}
K.~Hennessy {\em et~al.}, \ifthenelse{\boolean{articletitles}}{\emph{{Readout firmware of the vertex locator for LHCb Run 3 and beyond}}, }{}\href{https://doi.org/10.1109/TNS.2021.3085018}{IEEE Trans.\ Nucl.\ Sci.\  \textbf{68} (2021) 2472}\relax
\mciteBstWouldAddEndPuncttrue
\mciteSetBstMidEndSepPunct{\mcitedefaultmidpunct}
{\mcitedefaultendpunct}{\mcitedefaultseppunct}\relax
\EndOfBibitem
\bibitem{Passaro:2842603}
D.~Passaro, \ifthenelse{\boolean{articletitles}}{\emph{{Real-time luminosity and detector monitoring using FPGAs at the LHCb experiment}}, }{} Master's thesis, University of Pisa, 2022.
\newblock \url{https://cds.cern.ch/record/2842603}\relax
\mciteBstWouldAddEndPuncttrue
\mciteSetBstMidEndSepPunct{\mcitedefaultmidpunct}
{\mcitedefaultendpunct}{\mcitedefaultseppunct}\relax
\EndOfBibitem
\bibitem{Coombs:2021}
G.~Coombs, {\em {Beam-gas imaging at the LHCb experiment}}, PhD thesis, University of Glasgow, 2021\relax
\mciteBstWouldAddEndPuncttrue
\mciteSetBstMidEndSepPunct{\mcitedefaultmidpunct}
{\mcitedefaultendpunct}{\mcitedefaultseppunct}\relax
\EndOfBibitem
\bibitem{FerroLuzzi:2005em}
M.~Ferro-Luzzi, \ifthenelse{\boolean{articletitles}}{\emph{{Proposal for an absolute luminosity determination in colliding beam experiments using vertex detection of beam-gas interactions}}, }{}\href{https://doi.org/10.1016/j.nima.2005.07.010}{Nucl.\ Instrum.\ Meth.\  \textbf{A553} (2005) 388}\relax
\mciteBstWouldAddEndPuncttrue
\mciteSetBstMidEndSepPunct{\mcitedefaultmidpunct}
{\mcitedefaultendpunct}{\mcitedefaultseppunct}\relax
\EndOfBibitem
\bibitem{Barschel:2014iua}
C.~Barschel, {\em {Precision luminosity measurement at LHCb with beam-gas imaging}}, PhD thesis, RWTH Aachen U., 2014\relax
\mciteBstWouldAddEndPuncttrue
\mciteSetBstMidEndSepPunct{\mcitedefaultmidpunct}
{\mcitedefaultendpunct}{\mcitedefaultseppunct}\relax
\EndOfBibitem
\bibitem{vanderMeer:296752}
S.~{van der Meer}, \ifthenelse{\boolean{articletitles}}{\emph{{Calibration of the effective beam height in the ISR}}, }{} \href{http://cdsweb.cern.ch/search?p=CERN-ISR-PO-68-31&f=reportnumber&action_search=Search&c=LHCb} {CERN-ISR-PO-68-31}, 1968\relax
\mciteBstWouldAddEndPuncttrue
\mciteSetBstMidEndSepPunct{\mcitedefaultmidpunct}
{\mcitedefaultendpunct}{\mcitedefaultseppunct}\relax
\EndOfBibitem
\bibitem{Balagura:2020fuo}
V.~Balagura, \ifthenelse{\boolean{articletitles}}{\emph{{Van der Meer scan luminosity measurement and beam\textendash{}beam correction}}, }{}\href{https://doi.org/10.1140/epjc/s10052-021-08837-y}{Eur.\ Phys.\ J.\  \textbf{C81} (2021) 26}, \href{http://arxiv.org/abs/2012.07752}{{\normalfont\ttfamily arXiv:2012.07752}}\relax
\mciteBstWouldAddEndPuncttrue
\mciteSetBstMidEndSepPunct{\mcitedefaultmidpunct}
{\mcitedefaultendpunct}{\mcitedefaultseppunct}\relax
\EndOfBibitem
\bibitem{Belohrad:1267400}
D.~Belohrad {\em et~al.},  \ifthenelse{\boolean{articletitles}}{\emph{{The LHC fast BCT system: A comparison of design parameters with initial performance}}}{}, \href{https://cds.cern.ch/record/1267400}{CERN-BE-2010-010}, CERN, Geneva, 2010\relax
\mciteBstWouldAddEndPuncttrue
\mciteSetBstMidEndSepPunct{\mcitedefaultmidpunct}
{\mcitedefaultendpunct}{\mcitedefaultseppunct}\relax
\EndOfBibitem
\bibitem{Anders:1427726}
G.~Anders {\em et~al.},  \ifthenelse{\boolean{articletitles}}{\emph{{Study of the relative LHC bunch populations for luminosity calibration}}}{}, \href{https://cds.cern.ch/record/1427726}{{CERN-ATS-Note-2012-028}}, {CERN}, 2012\relax
\mciteBstWouldAddEndPuncttrue
\mciteSetBstMidEndSepPunct{\mcitedefaultmidpunct}
{\mcitedefaultendpunct}{\mcitedefaultseppunct}\relax
\EndOfBibitem
\bibitem{Barschel:1425904}
C.~Barschel {\em et~al.}, \ifthenelse{\boolean{articletitles}}{\emph{{Results of the LHC DCCT calibration studies}}, }{} \href{http://cdsweb.cern.ch/search?p=CERN-ATS-Note-2012-026&f=reportnumber&action_search=Search&c=LHCb} {CERN-ATS-Note-2012-026}, 2012\relax
\mciteBstWouldAddEndPuncttrue
\mciteSetBstMidEndSepPunct{\mcitedefaultmidpunct}
{\mcitedefaultendpunct}{\mcitedefaultseppunct}\relax
\EndOfBibitem
\bibitem{Babaev:2023fim}
A.~Babaev {\em et~al.}, \ifthenelse{\boolean{articletitles}}{\emph{{Impact of beam{\textendash}beam effects on absolute luminosity calibrations at the CERN Large Hadron Collider}}, }{}\href{https://doi.org/10.1140/epjc/s10052-023-12192-5}{Eur.\ Phys.\ J.\  \textbf{C84} (2024) 17}, \href{http://arxiv.org/abs/2306.10394}{{\normalfont\ttfamily arXiv:2306.10394}}\relax
\mciteBstWouldAddEndPuncttrue
\mciteSetBstMidEndSepPunct{\mcitedefaultmidpunct}
{\mcitedefaultendpunct}{\mcitedefaultseppunct}\relax
\EndOfBibitem
\bibitem{Emittance.PhysRevAccelBeams}
M.~Hostettler {\em et~al.}, \ifthenelse{\boolean{articletitles}}{\emph{{Luminosity scans for beam diagnostics}}, }{}\href{https://doi.org/10.1103/PhysRevAccelBeams.21.102801}{Phys.\ Rev.\ Accel.\ Beams \textbf{21} (2018) 102801}\relax
\mciteBstWouldAddEndPuncttrue
\mciteSetBstMidEndSepPunct{\mcitedefaultmidpunct}
{\mcitedefaultendpunct}{\mcitedefaultseppunct}\relax
\EndOfBibitem
\bibitem{Bertulani_2005}
C.~A. Bertulani, S.~R. Klein, and J.~Nystrand, \ifthenelse{\boolean{articletitles}}{\emph{{Physics of ultra-peripheral nuclear collisions}}, }{}\href{https://doi.org/10.1146/annurev.nucl.55.090704.151526}{Annu.\ Rev.\ Nucl.\ Part.\ Sci.\  \textbf{55} (2005) 271–310}\relax
\mciteBstWouldAddEndPuncttrue
\mciteSetBstMidEndSepPunct{\mcitedefaultmidpunct}
{\mcitedefaultendpunct}{\mcitedefaultseppunct}\relax
\EndOfBibitem
\bibitem{Rinnert:2015uns}
K.~Rinnert, \ifthenelse{\boolean{articletitles}}{\emph{{LHCb silicon detectors: the Run 1 to Run 2 transition and first experience of Run 2}}, }{}\href{https://doi.org/10.22323/1.254.0004}{PoS \textbf{VERTEX2015} (2015) 004}\relax
\mciteBstWouldAddEndPuncttrue
\mciteSetBstMidEndSepPunct{\mcitedefaultmidpunct}
{\mcitedefaultendpunct}{\mcitedefaultseppunct}\relax
\EndOfBibitem
\bibitem{cordova2025}
G.~Cordova {\em et~al.}, \ifthenelse{\boolean{articletitles}}{\emph{{Real-time monitoring of LHCb interaction region with a fast trackless methodology}}, }{}\href{http://arxiv.org/abs/2503.10831}{{\normalfont\ttfamily arXiv:2503.10831}}\relax
\mciteBstWouldAddEndPuncttrue
\mciteSetBstMidEndSepPunct{\mcitedefaultmidpunct}
{\mcitedefaultendpunct}{\mcitedefaultseppunct}\relax
\EndOfBibitem
\end{mcitethebibliography}

\newpage
\centerline
{\large\bf LHCb collaboration}
\begin
{flushleft}
\small
R.~Aaij$^{38}$\lhcborcid{0000-0003-0533-1952},
A.S.W.~Abdelmotteleb$^{57}$\lhcborcid{0000-0001-7905-0542},
C.~Abellan~Beteta$^{51}$\lhcborcid{0009-0009-0869-6798},
F.~Abudin\'en$^{57}$\lhcborcid{0000-0002-6737-3528},
T.~Ackernley$^{61}$\lhcborcid{0000-0002-5951-3498},
A. A. ~Adefisoye$^{69}$\lhcborcid{0000-0003-2448-1550},
B.~Adeva$^{47}$\lhcborcid{0000-0001-9756-3712},
M.~Adinolfi$^{55}$\lhcborcid{0000-0002-1326-1264},
P.~Adlarson$^{85}$\lhcborcid{0000-0001-6280-3851},
C.~Agapopoulou$^{14}$\lhcborcid{0000-0002-2368-0147}
% ...

{\footnotesize \it

% ...
$^{14}$Universit{\'e} Paris-Saclay, CNRS/IN2P3, IJCLab, Orsay, France\\
% ...
$^{38}$Nikhef National Institute for Subatomic Physics, Amsterdam, Netherlands\\
% ...
$^{47}$Instituto Galego de F{\'\i}sica de Altas Enerx{\'\i}as (IGFAE), Universidade de Santiago de Compostela, Santiago de Compostela, Spain\\
% ...
$^{51}$Physik-Institut, Universit{\"a}t Z{\"u}rich, Z{\"u}rich, Switzerland\\
% ...
$^{55}$H.H. Wills Physics Laboratory, University of Bristol, Bristol, United Kingdom\\
% ...
$^{57}$Department of Physics, University of Warwick, Coventry, United Kingdom\\
% ...
$^{60}$School of Physics and Astronomy, University of Glasgow, Glasgow, United Kingdom\\
$^{61}$Oliver Lodge Laboratory, University of Liverpool, Liverpool, United Kingdom\\
% ...
$^{69}$Syracuse University, Syracuse, NY, United States\\
% ...
$^{85}$Department of Physics and Astronomy, Uppsala University, Uppsala, Sweden, associated to $^{60}$\\
% ...
}
\end{flushleft}

\end{document}